\documentclass[twocolumn,numberedappendix]{AASTeX62}
\usepackage{hyperref}
\usepackage{mathtools}
\usepackage{footnote}
\usepackage{todonotes}

\submitjournal{ApJ}

\shorttitle{Comparing Redundant and sky-based calibration}
\shortauthors{Li et al.}

\begin{document}

\title{Comparing Redundant and Sky Model Based Interferometric Calibration: A First Look with Phase II of the MWA}
\correspondingauthor{W.~Li}
\email{wenyang\_li@brown.edu}
\author{W.~Li}
\affiliation{Department of Physics, Brown University, Providence, RI 02912, USA}
\author{J.~C.~Pober}
\affiliation{Department of Physics, Brown University, Providence, RI 02912, USA}
\author{B.~J.~Hazelton}
\affiliation{Department of Physics, University of Washington, Seattle, WA 98195, USA}
\author{N.~Barry}
\affiliation{Department of Physics, University of Washington, Seattle, WA 98195, USA}
\author{M.~F.~Morales}
\affiliation{Department of Physics, University of Washington, Seattle, WA 98195, USA}
\author{I.~Sullivan}
\affiliation{Department of Physics, University of Washington, Seattle, WA 98195, USA}
\author{A.~R.~Parsons}
\affiliation{University of California, Berkeley, Astronomy Dept., 501 Campbell Hall \#3411, Berkeley, CA 94720, USA}
\author{Z.~S.~Ali}
\affiliation{University of California, Berkeley, Astronomy Dept., 501 Campbell Hall \#3411, Berkeley, CA 94720, USA}
\author{J.~S.~Dillon}
\affiliation{University of California, Berkeley, Astronomy Dept., 501 Campbell Hall \#3411, Berkeley, CA 94720, USA}
\author{A.~P.~Beardsley}
\affiliation{School of Earth and Space Exploration, Arizona  State University, Tempe, AZ 85287, USA}
\author{J.~D.~Bowman}
\affiliation{School of Earth and Space Exploration, Arizona  State University, Tempe, AZ 85287, USA}
\author{F.~Briggs}
\affiliation{Research School of Astronomy and Astrophysics, Australian National University, Canberra, ACT 2611, Australia}
\author{R.~Byrne}
\affiliation{Department of Physics, University of Washington, Seattle, WA 98195, USA}
\author{P.~Carroll}
\affiliation{Department of Physics, University of Washington, Seattle, WA 98195, USA}
\author{B.~Crosse}
\affiliation{International Centre for Radio Astronomy Research, Curtin University, Bentley, WA 6102, Australia}
\author{D.~Emrich}
\affiliation{International Centre for Radio Astronomy Research, Curtin University, Bentley, WA 6102, Australia}
\author{A.~Ewall-Wice}
\affiliation{MIT Kavli Institute for Astrophysics and Space  Research, Massachusetts Institute of Technology, Cambridge, MA  02139, USA}
\author{L.~Feng}
\affiliation{MIT Kavli Institute for Astrophysics and Space  Research, Massachusetts Institute of Technology, Cambridge, MA  02139, USA}
\author{T.~M.~O.~Franzen}
\affiliation{International Centre for Radio Astronomy Research, Curtin University, Bentley, WA 6102, Australia}
\author{J.~N.~Hewitt}
\affiliation{MIT Kavli Institute for Astrophysics and Space  Research, Massachusetts Institute of Technology, Cambridge, MA  02139, USA}
\author{L.~Horsley}
\affiliation{International Centre for Radio Astronomy Research, Curtin University, Bentley, WA 6102, Australia}
\author{D.~C.~Jacobs}
\affiliation{School of Earth and Space Exploration, Arizona  State University, Tempe, AZ 85287, USA}
\author{M.~Johnston-Hollitt}
\affiliation{International Centre for Radio Astronomy Research, Curtin University, Bentley, WA 6102, Australia}
\affiliation{Peripety Scientific Ltd., PO Box 11355 Manners Street, 6142 Wellington, New Zealand}
\author{C.~Jordan}
\affiliation{International Centre for Radio Astronomy Research, Curtin University, Bentley, WA 6102, Australia}
\author{R.~C.~Joseph}
\affiliation{ICRAR University of Western Australia, Crawley, WA 6009, Australia}
\affiliation{International Centre for Radio Astronomy Research, Curtin University, Bentley, WA 6102, Australia}
\affiliation{ARC Centre of Excellence for All-sky  Astrophysics (CAASTRO)}
\author{D.~L.~Kaplan}
\affiliation{Department of Physics, University of Wisconsin--Milwaukee, Milwaukee, WI 53201, USA}
\author{D.~Kenney}
\affiliation{International Centre for Radio Astronomy Research, Curtin University, Bentley, WA 6102, Australia}
\author{H.~Kim}
\affiliation{School of Physics, The University of Melbourne, Parkville, VIC 3010, Australia}
\author{P.~Kittiwisit}
\affiliation{School of Earth and Space Exploration, Arizona  State University, Tempe, AZ 85287, USA}
\author{A.~Lanman}
\affiliation{Department of Physics, Brown University, Providence, RI 02912, USA}
\author{J.~Line}
\affiliation{School of Physics, The University of Melbourne, Parkville, VIC 3010, Australia}
\author{B.~McKinley}
\affiliation{School of Physics, The University of Melbourne, Parkville, VIC 3010, Australia}
\author{D.~A.~Mitchell}
\affiliation{CSIRO Astronomy and Space Science (CASS), PO Box  76, Epping, NSW 1710, Australia}
\affiliation{ARC Centre of Excellence for All-sky  Astrophysics (CAASTRO)}
\author{S.~Murray}
\affiliation{International Centre for Radio Astronomy Research, Curtin University, Bentley, WA 6102, Australia}
\author{A.~Neben}
\affiliation{MIT Kavli Institute for Astrophysics and Space  Research, Massachusetts Institute of Technology, Cambridge, MA  02139, USA}
\author{A.~R.~Offringa}
\affiliation{ASTRON, The Netherlands Institute for Radio Astronomy, Postbus 2, 7990 AA, Dwingeloo, The Netherlands}
\author{D.~Pallot}
\affiliation{ICRAR University of Western Australia, Crawley, WA 6009, Australia}
\author{S.~Paul}
\affiliation{Raman Research Institute, Bangalore 560080, India}
\author{B.~Pindor}
\affiliation{School of Physics, The University of Melbourne, Parkville, VIC 3010, Australia}
\author{P.~Procopio}
\affiliation{School of Physics, The University of Melbourne, Parkville, VIC 3010, Australia}
\author{M.~Rahimi}
\affiliation{School of Physics, The University of Melbourne, Parkville, VIC 3010, Australia}
\affiliation{ARC  Centre  of  Excellence  for  All  Sky  Astrophysics  in  3  Dimensions  (ASTRO  3D)}
\author{J.~Riding}
\affiliation{School of Physics, The University of Melbourne, Parkville, VIC 3010, Australia}
\author{S.~K.~Sethi}
\affiliation{Raman Research Institute, Bangalore 560080, India}
\author{N.~Udaya~Shankar}
\affiliation{Raman Research Institute, Bangalore 560080, India}
\author{K.~Steele}
\affiliation{International Centre for Radio Astronomy Research, Curtin University, Bentley, WA 6102, Australia}
\author{R.~Subrahmanian}
\affiliation{Raman Research Institute, Bangalore 560080, India}
\author{M.~Tegmark}
\affiliation{MIT Kavli Institute for Astrophysics and Space  Research, Massachusetts Institute of Technology, Cambridge, MA  02139, USA}
\author{N.~Thyagarajan}
\affiliation{National Radio Astronomy Observatory, 1003 Lopezville Rd., Socorro, NM 87801, USA}
\affiliation{School of Earth and Space Exploration, Arizona  State University, Tempe, AZ 85287, USA}
\affiliation{Jansky Fellow of the National Radio Astronomy Observatory}
\author{S.~J.~Tingay}
\affiliation{International Centre for Radio Astronomy Research, Curtin University, Bentley, WA 6102, Australia}
\affiliation{ARC Centre of Excellence for All-sky  Astrophysics (CAASTRO)}
\affiliation{Osservatorio di Radio Astronomia, Istituto Nazionale di Astrofisica, Bologna, Italy, 40123}
\author{C.~Trott}
\affiliation{International Centre for Radio Astronomy Research, Curtin University, Bentley, WA 6102, Australia}
\affiliation{ARC Centre of Excellence for All-sky  Astrophysics (CAASTRO)}
\author{M.~Walker}
\affiliation{International Centre for Radio Astronomy Research, Curtin University, Bentley, WA 6102, Australia}
\author{R.~B.~Wayth}
\affiliation{International Centre for Radio Astronomy Research, Curtin University, Bentley, WA 6102, Australia}
\affiliation{ARC Centre of Excellence for All-sky  Astrophysics (CAASTRO)}
\author{R.~L.~Webster}
\affiliation{School of Physics, The University of Melbourne, Parkville, VIC 3010, Australia}
\affiliation{ARC Centre of Excellence for All-sky  Astrophysics (CAASTRO)}
\affiliation{ARC  Centre  of  Excellence  for  All  Sky  Astrophysics  in  3  Dimensions  (ASTRO  3D)}
\author{A.~Williams}
\affiliation{International Centre for Radio Astronomy Research, Curtin University, Bentley, WA 6102, Australia}
\author{C.~Wu}
\affiliation{ICRAR University of Western Australia, Crawley, WA 6009, Australia}
\author{S.~Wyithe}
\affiliation{School of Physics, The University of Melbourne, Parkville, VIC 3010, Australia}

\begin{abstract}

Interferometric arrays seeking to measure the 21\,cm signal from the Epoch of Reionization must contend with overwhelmingly bright emission from foreground sources. Accurate recovery of the 21\,cm signal will require precise calibration of the array, and several new avenues for calibration have been pursued in recent years, including methods using redundancy in the antenna configuration. The newly upgraded Phase II of Murchison Widefield Array (MWA) is the first interferometer that has large numbers of redundant baselines while retaining good instantaneous UV-coverage. This array therefore provides a unique opportunity to compare redundant calibration with sky-model based algorithms. In this paper, we present the first results from comparing both calibration approaches with MWA Phase II observations. For redundant calibration, we use the package \texttt{OMNICAL}, and produce sky-based calibration solutions with the analysis package Fast Holographic Deconvolution (FHD). There are three principal results.  (1) We report the success of \texttt{OMNICAL} on observations of ORBComm satellites, showing substantial agreement between redundant visibility measurements after calibration. (2) We directly compare \texttt{OMNICAL} calibration solutions with those from FHD, and demonstrate these two different calibration schemes give extremely similar results. (3) We explore improved calibration by combining \texttt{OMNICAL} and FHD. We evaluate these combined methods using power spectrum techniques developed for EoR analysis and find evidence for marginal improvements mitigating artifacts in the power spectrum.  These results are likely limited by signal-to-noise in the six hours of data used, but suggest future directions for combining these two calibration schemes.

\end{abstract}


\section{Introduction}

21~cm observations of the Epoch of Reionization (EoR) have the potential to reveal a wealth of information about the formation of the first stars and galaxies by measuring the three dimensional power spectrum and full tomographic maps of the neutral intergalactic medium (IGM; \citealt{morales2010reionization, furlanetto201621}). However, these observations are technically very challenging due to bright astrophysical foregrounds, the complex frequency dependence of instrumental response of radio interferometers, radio frequency interference (RFI), and the effects of the ionosphere.

Recent work has highlighted the critical role precision instrument calibration will play in disentangling the faint cosmological signal from the bright foregrounds \citep{Barry2016,ewall2017impact,trott2016spectral, patil2016systematic}. Current calibration efforts for EoR observations largely fall into two camps:  sky-based calibration using deep foreground catalogs and forward modeling of the instrument visibilities \citep{Beardsley2016,dillon2015empirical,trott2016chips,patil2016systematic,Carroll2016,procopio2017high,Hurley-Walker2016,intema2017gmrt}, and redundant calibration that foregoes a sky model but requires the antennas be placed on a regular grid \citep{Wieringa1992,Liu2010,Zheng2014}. 

To date it has been impossible to directly compare the efficacy of the two calibration approaches on real data. Redundant arrays tend to have very poor UV-coverage, and are thus hard to calibrate with sky-based approaches \citep{parsons2012sensitivity,Zheng2016}, and arrays with good imaging performance do not have the regular antenna layout necessary for redundant calibration.

Using new observations with Phase II \citep{MWAPhaseII} of the Murchison Widefield Array (MWA; \citealt{Tingay2013,bowman2013mwa})  we report on the first direct comparison of sky and redundant calibration with an Epoch of Reionization instrument. During Phase I, the MWA consisted of 128 antenna tiles in a pseudo-random layout designed for excellent instantaneous uv coverage.  Phase II added 128 additional tiles (for a total of 256), but only 128 can be correlated simultaneously.  Phase II therefore operates in two modes: a compact array and an extended array, each consisting of a subset of the 256 total available tiles. 
In the compact array new tiles were added in two hexagonal cores (see Section \ref{obsSec}), providing a hybrid data set with both redundant baselines and the excellent imaging characteristics of the existing MWA array \citep{beardsley2012new}. We use data from this unique array to directly compare redundant and sky-based calibration.

The structure of the remainder of this paper is as follows.
In Section \ref{obsSec}, we further describe the compact array of Phase II of the MWA and the observations used in our analysis.
In Section \ref{calSec}, we describe the calibration techniques used to perform both sky-based and redundancy-based calibration.
We also develop and present new tools needed to map between the calibration approaches (Section \ref{DPSec}). 
In Section \ref{orbcommSec}, we present the results of applying redundant calibration to observations of the ORBComm satellite system, and in Section \ref{comparisonSec}, we directly compare sky-based and redundant calibration solutions derived from observations of an EoR target field.
In Section \ref{combiningSec}, we 
explore ways of combining the calibration results and compare the resulting EoR power spectra (PS).
We discuss potential shortcomings of our analysis in Section \ref{discussionSec}, and conclude in Section \ref{conclusionSec}.

\section{Observations}
\label{obsSec}
\subsection{Phase II of the MWA}
\begin{figure}
\centering
\includegraphics[width=\linewidth]{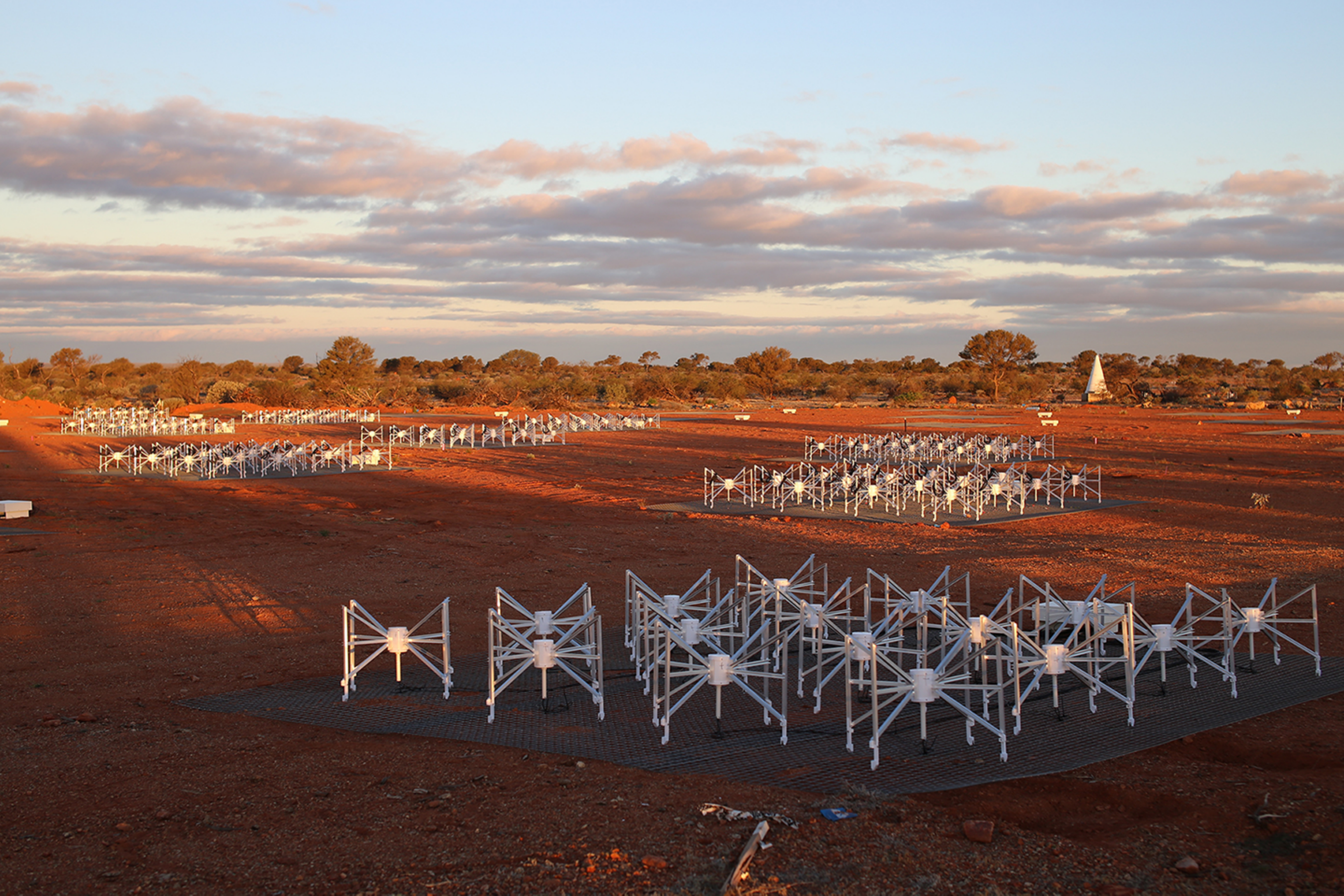}
\caption{MWA Phase II tiles in the Murchison Radio-Astronomy Observatory in Western Australia. (Taken by Greg Rowbotham in June~2016 when Phase II was under construction.)}
\label{FIGtiles}
\end{figure}

MWA Phase II compact array consists of 128 tiles. 
Each tile includes 16 dual-polarization dipoles, as shown in Figure \ref{FIGtiles}. 72 of the tiles are configured into two hexagons with high redundancy for redundant calibratability and power spectrum sensitivity. The other 56 tiles are arranged with minimal redundancy; 8 of these tiles are located at two to three hundred meters from the core to provide extended baselines for better imaging and survey capabilities. 

\begin{figure}
    \centering
    \includegraphics[width=\linewidth]{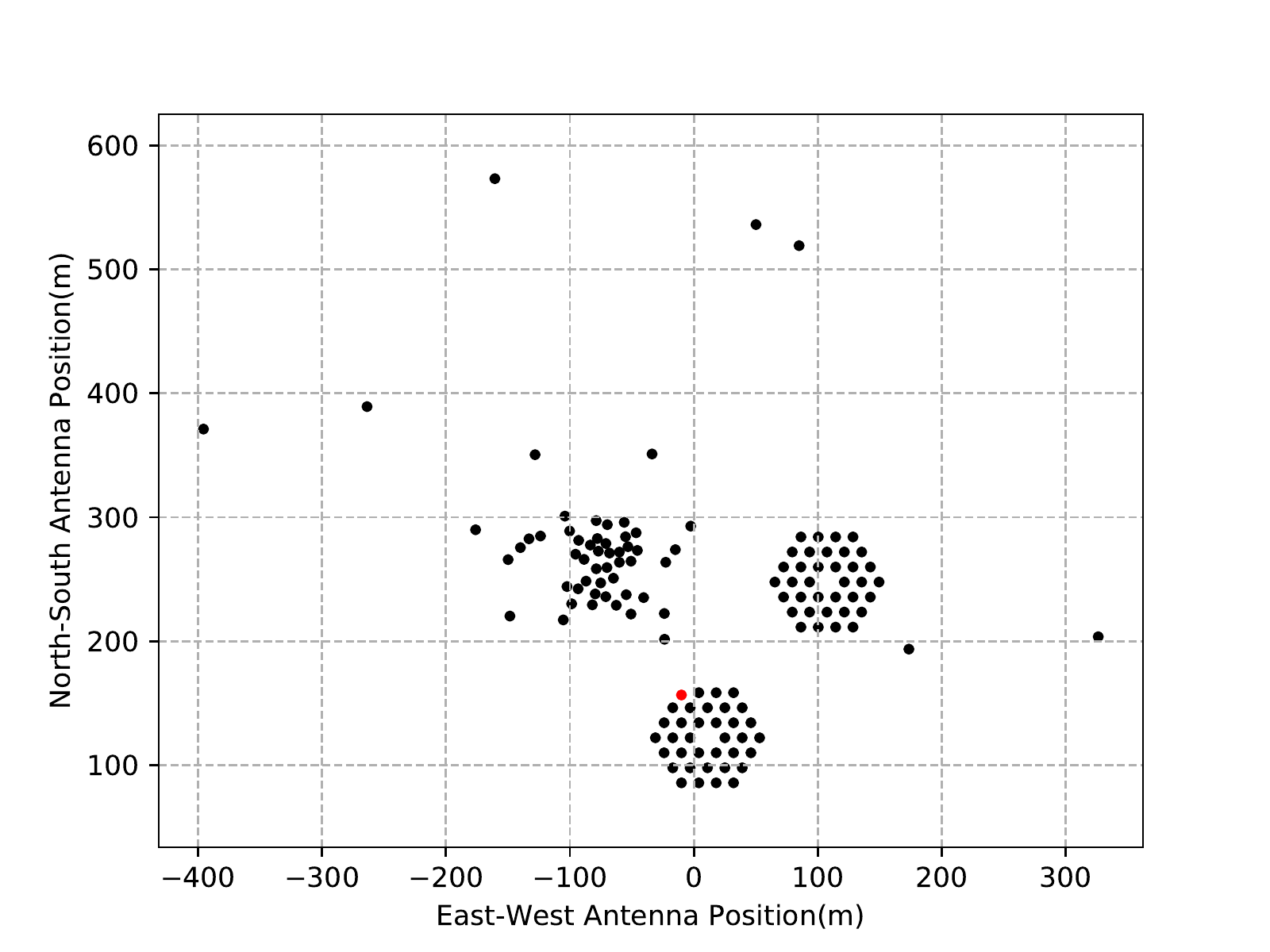}
    \caption{Top: MWA Phase II Configuration. Tile 1037 (red) on the upper left corner of the south hexagon was flagged because the site terrain prevented its placement. Bottom: Tile positions of the north hexagon, with tile numbers labeled.}
    \label{FIGpos}
\end{figure}

The upper panel of Figure \ref{FIGpos} shows the configuration of all 128 tiles of MWA Phase II; the lower panel shows the north hexagon, with tile numbers labeled. All tiles in the north hexagon are labeled from 1001 to 1036 (bottom plot in Figure \ref{FIGpos}), and tiles in the south hexagon are labeled from 1037 to 1072. Due to ground conditions at the MWA site, one of the tiles in the south hexagon (tile 1037) could not be placed at the position where the corner of the hexagon should be so it is flagged, leaving 71 hexagon tiles and 56 non-hexagon tiles. The hexagon shaped configuration is designed for two reasons: increased sensitivity on short baselines for power spectrum measurements \citep{parsons2012sensitivity} and opportunities for redundant calibration.

\subsection{The Data}
The data we processed in this work are from MWA Phase II compact array observations of the EoR0 field ($RA=0^{\circ}$, $Dec=-27^{\circ}$)  at frequencies of $167-197$\, MHz (corresponding to a 21\, cm redshift of $7.5-6.2$).  Observations were taken on 2016 November 17, from 11:26 to 13:10, 2016 November 19, from 11:18 to 13:02, and 2016 November 21, from 11:19 to 12:54 (UTC), as well as 2 minutes of ORBComm satellite observations at $134-164$\, MHz on 2016 September 21 at 18:43 (UTC). The total band was divided into 24 1.28 MHz sub-bands; each sub-band is further sub-divided into 32 fine channels with a frequency resolution of 40 kHz. The time of observation per data file was 112 seconds, with a time resolution of 0.5 seconds. The data was preprocessed by the COTTER pipeline \citep{Offringa2015}, which uses AOFlagger\footnote{\url{http://aoflagger.sourceforge.net/doc/api/}} to flag radio frequency interference (RFI), reduces data volume by averaging in time and frequency, and converts data into the \texttt{uvfits} format. In this work, we average the EoR0 data to 2 second time integrations and 80 kHz frequency resolution. The ORBComm data were averaged into 4 second time integrations and 40 kHz frequency resolution. We choose the EoR0 high band observation because this is one of the best-studied fields with the MWA \citep{Carroll2016,Beardsley2016}. It has low sky temperature and relatively few bright, resolved sources which leads to better EoR sensitivity \citep{Jacobs2016}. The ORBComm observation is for testing redundant calibration because of its high signal to noise. 

In the MWA, an analog beamformer can steer the main lobe of the tile primary beam to change the field being observed. For EoR observerations, we use the ``drift-and-shift" method, where we observe specific pointings with the beamformer and allow the sky to drift overhead for some duration before re-pointing. The EoR0 observations we used in this work include 5 pointings, and each pointing spans 30 minutes.\footnote{The pointings are labeled as -2, -1, 0, 1, and 2, where 0 corresponds to a zenith pointing. Pointing -2 and 2 have less data, i.e., less than 30 minutes.} The 2 minutes ORBComm Observation consists of a single pointing towards zenith.\\

\section{Calibration Techniques}

\label{calSec}
Discrepancies between measured data visibilities and true visibilities can have different causes: instrumental gains, cross talk between tiles, RFI, thermal noise, tile pointing error, ionosphere distortion, etc. In this work, we only consider the contribution from the analog/digital electronics of each tile and mainly focus on the complex antenna-based instrumental gain calibration. In this section, we will briefly show the basic mathematical background of both sky calibration and redundant calibration and describe the specific software packages we use to perform them.

\subsection{Assumptions}
The instrumental calibration is assumed to be tile-based. At a given polarization $p$, given frequency channel $\nu$, and given time step $t$, the basic assumption of the relation between the measured visibility $v_{ij}$ recorded by the baseline $ij$ (the baseline formed by tile $i$ and tile $j$) and the true visibility $y_{ij}$ is described by Equation \ref{EQassumption}, where $g_i$ and $g_j$ are the complex gains of tile $i$ and tile $j$, respectively, and $n_{ij}$ is a random noise term. 

\begin{equation}
v_{ij}(t,\nu,p)\approx g_i(t,\nu,p)g_j^*(t,\nu,p)y_{ij}(t,\nu,p)+n_{ij}
\label{EQassumption}
\end{equation}

In the case of MWA, the tile gains vary from pointing to pointing due to the change in tile beams. As we observed from real data calibration (using both sky-based calibration and redundant calibration), gains of the same pointing also vary from day to day, but are relatively stable over time within one pointing (30 minutes). \cite{barrythesis} has demonstrated that the gain amplitudes are stable if the ambient temperature does not change. Therefore, we assume that 30 minutes of a single pointing is the longest time scale within which we can consider the instrumental gains to be time independent.

Our goal is to solve for the gain per time, per frequency channel, per polarization for each tile using two different methodologies: 1. generate model visibilities based on the combination of our best models for the sky, array layout, and tile primary beam, then minimize the difference between model visibilities and data (\emph{sky calibration}); 2. using redundancy, minimize the differences among the measurements from redundant baselines (\emph{redundant calibration}).

\subsection{FHD sky-based calibration}
\label{fhdcalSec}
Fast Holographic Deconvolution (FHD\footnote{https://github.com/EoRImaging/FHD} \citealt{Sullivan2012}) is a software package which provides interferometric data simulation, calibration, and imaging. In this paper, we will use the FHD framework as our method to do sky calibration.

In Equation \ref{EQassumption}, the true visibility $y_{ij}$ consists of foregrounds and EoR signal. We neglect the EoR term because it is orders of magnitude smaller than the foreground term. If we have reasonable knowledge of the foreground sources, we can generate model visibilities $m_{ij}$, and replace $y_{ij}$ with $m_{ij}$ in Equation \ref{EQassumption}, as Equation \ref{EQfhdasum} shows: 

\begin{equation}
v_{ij}\approx g_ig_j^*m_{ij}+n_{ij}
\label{EQfhdasum}
\end{equation}
We use a sky model developed by \cite{Carroll2016} specifically for the EoR0 field, which contains about 11000 point sources in the field of view. 
We then solve for the gains by evaluating $\chi^2$ in Equation \ref{EQfhdchi}, making it a least squares problem, with $2\times N_{tiles}-1$\footnote{The gains are complex so the number is multiplied by 2; The overall absolute phase parameter is constrained by picking a phase reference tile.} parameters to solve:
\begin{equation}
\chi^2=\sum_{ij}\frac{|v_{ij}-g_ig_j^*m_{ij}|^2}{\sigma_{ij}^2}
\label{EQfhdchi}
\end{equation}
$\sigma_{ij}^2$ is the noise variance of baseline $ij$\footnote{In FHD framework, the noise does not contribute to linear least squares solver, i.e., $\sigma_{ij}\equiv 1$, assuming all baselines having the same noise level.}. We solve for each $g_i$ per polarization per frequency channel by feeding an initial guess of the gain solutions (generally all ones by default), fixing all other $g_j$($j\neq{i}$), minimizing the $\chi^2$ to get a new guess for $g_i$, then average it with the previous guess of $g_i$; this average is treated as the solution for $g_i$, and we then run the previous process iteratively until the solutions converge \citep{Salvini2014}.

Following this per-tile, per-frequency, per-polarization sky-based calibration, FHD reduces the number of calibration parameters by computing an average bandpass over subsets of the tiles and then only allowing tile-to-tile deviations from this average solution to be smooth in frequency \citep{Beardsley2016,Barry2016}.
The exact form of the final calibration solutions $g_i(\nu)$ for tile $i$ is given by:
\begin{equation}
\label{EQbpfit}
g_i(\nu) = B_c(\nu)[(\alpha_{0,i}+\alpha_{1,i}\nu+\alpha_{2,i}\nu^2)e^{2\pi i(\beta_{0,i}+\beta_{1,i}\nu)}+R_i(\nu)]
\end{equation}
$B_c(\nu)$ is a tile-independent bandpass amplitude calculated by averaging the amplitude gains over all tiles which share a cable type. In the MWA Phase II design, each tile has one of 4 distinct lengths of cable leading from its beamformer to the receiver: 90 m, 150 m, 230 m, or 320 m.  This design leads to 4 subtly different bandpass responses $B_c(\nu)$'s due to different filters used on different cable types and imperfect terminations \citep{Beardsley2016,barrythesis}.
For each tile, deviations from the per-cable type bandpass $B_c(\nu)$ are fit with low order polynomials in frequency, with coefficients $\alpha_{0,i},\ \alpha_{1,i},\ \alpha_{2,i},\ \beta_{0,i},\ \beta_{1,i}$ ($\alpha$'s for the amplitude and $\beta$'s for the phase). The final parameter, $R_i(\nu)$, is the strongest sinusoidal cable reflection mode found for tile $i$, which is a complex number. In this work, we only fit $R_i(\nu)$ for 150 m cables which have the strongest reflection \citep{Barry2016,Beardsley2016}. 
The motivation for this fitting methodology is to mitigate frequency-dependent errors introduced by an incomplete sky model, which can lead to foreground contamination of an EoR signal \citep{Barry2016,Beardsley2016}.

\subsection{Redundant Calibration (\texttt{OMNICAL})}

Mathematically, redundant calibration requires sufficient baselines to measure the same Fourier mode of the sky emission so that the there are more measurements than the number of unknown visibilities and tile gains \citep{Liu2010}. In Phase II data, only the two hexagonal sub-arrays are redundantly calibratable. For one time step, one frequency channel, and one polarization, the unknown parameters consist of tile gains (for those tiles that participate in a minimum number of redundant baselines) and the visibilities themselves for each unique type of baseline. If only redundant baseline groups containing at least 2 baselines are considered, there are 71 tiles and 181 unique baseline types and therefore 252 free parameters to fit, while the number of measurements is 2477.\footnote{All parameters are complex numbers, thus the number of fitted parameters, as well as the number of measurements, are multiplied by two.}

In this paper, we use the package \texttt{OMNICAL}\footnote{https://github.com/jeffzhen/omnical} \citep{Zheng2014} for redundant calibration. When running \texttt{OMNICAL} on the data, we load \texttt{uvfits} data files using the open source python module \texttt{pyuvdata}\footnote{https://github.com/HERA-Team/pyuvdata}\citep{JHazelton2017}.

\texttt{OMNICAL} consists of two algorithms: a logarithmic method (\texttt{logcal}) and a linearized method (\texttt{lincal} \citealt{Liu2010,Zheng2014}). To interpret the algorithms, we express the gain in the form of $g_i=e^{\eta_i+i\phi_i}$ with $e^{\eta_i}$ the amplitude and $\phi_i$ the phase of tile $i$.

In \texttt{logcal}, we linearize the equations by taking the logarithm of Equation \ref{EQassumption}, where the noise contribution is represented by $\omega_{ij}=\ln(1+\frac{n_{ij}}{g_ig_j^*y_{ij}})$. This gives
\begin{equation}
\ln(v_{ij})=\eta_i+\eta_j+i(\phi_i-\phi_j)+\ln(y_{ij})+\omega_{ij}
\label{EQlogcal}
\end{equation}
By separating the real and imaginary parts of Equation \ref{EQlogcal}, the amplitude terms and phase terms are separated. We solve for the gains by minimizing Equation \ref{EQamp} and \ref{EQphs}, which are the linear least squares equations for the amplitudes and phases, respectively.
\begin{equation}
\sum_{ij}[\ln|v_{ij}|-\eta_i-\eta_j-\ln|y_{ij}|]^2
\label{EQamp}
\end{equation}
\begin{equation}
\sum_{ij}[\arg(v_{ij})-\phi_i+\phi_j-\arg(y_{ij})]^2
\label{EQphs}
\end{equation}
However, the \texttt{logcal} method is biased. The noise is assumed to be Gaussian and to have zero mean in real/imaginary space, while this is not the case in amplitude/phase space \citep{Liu2010}. To address this issue, \texttt{lincal} is introduced.  

In \texttt{lincal}, we perform a Taylor expansion on Equation \ref{EQassumption} about some fiducial guess $g_i^0$'s for the gains and $y_{ij}^0$'s for the true visibilities, which leads to Equation \ref{EQlincal}.

\begin{equation}
v_{ij} \approx g_i^0{g_j^0}^*y_{ij}^0 + {g_j^0}^*y_{ij}^0\Delta g_i+g_i^0y_{ij}^0\Delta g_j^*+g_i^0{g_j^0}^*\Delta y_{ij}
 \label{EQlincal}
\end{equation}
where $\Delta g_i = g_i - g_i^0$ and $\Delta{y_{ij}}=y_{ij}-y_{ij}^0$. This expansion linearizes Equation \ref{EQassumption} so that we can employ a least-squares fit to solve for the $\Delta g_i$'s, and $\Delta{y_{ij}}$'s.  The initial fiducial guess is required to be in a local minima around the true solution; we use the \texttt{logcal} solutions as the initial guesses for \texttt{lincal}. After we have the solutions for $g_i$ and $y_{ij}$, we take them as our new fiducial guess and feed them into \texttt{lincal}, and run this process iteratively. \texttt{lincal} solves in real and imaginary space, so if the noise level for all baselines is the same, the least-squares fit is unbiased.\\

Before we start the calibration, we have to deal with a phase wrapping problem: there is ambiguity between 0 and $2\pi$ in phase. For example, the difference between a phase of $359^{\circ}$ and $1^{\circ}$ is $358^{\circ}$ instead of $2^{\circ}$. If there is no pre-calibration before \texttt{logcal}, these calibration procedures can potentially take a small difference in phase and drive it in the opposite direction instead of further minimizing it. As a result, the calibration is not handled properly and the solutions do not converge.
To overcome phase wrapping, we introduce \texttt{firstcal} as our pre-calibration method, i.e., to get an initial estimate of phase solutions.

\subsubsection{Firstcal method}

We use the \texttt{firstcal} module developed by the HERA team to find a per tile delay to provide an initial phase solution\footnote{\url{https://github.com/HERA-Team/hera_cal}} using array redundancy, without any reference to the sky. \texttt{firstcal} takes visibility pairs $v_{ij}$ and $v_{kl}$ from the same redundant baseline group, calculates the product of $v_{ij}$ and $v_{kl}^*$. If complex gains of all four tiles differed only by a single per-tile delay, $\tau_i$, then the time-average of this quantity, $R_{ijkl}$, is given by
\begin{equation}
\begin{split}
R_{ijkl}(\nu)=& \langle v_{ij}(\nu,t)v_{kl}^{*}(\nu,t)\rangle_t \\
=&A^2(\nu)\exp(2{\pi}i\nu(\tau_i-\tau_j-\tau_k+\tau_l)).
\end{split}
\label{EQrphs}
\end{equation}
Here $\nu$ is frequency and $A$ is visibility amplitude. Multiplying $v_{ij}$ by $v_{kl}^*$ cancels out the frequency structure of the visibilities, leaving only the exponential of the four tile delays. The Fourier transform of Equation \ref{EQrphs} along the frequency axis (i.e. the delay transform; \citealt{Parsons2009}), should be peaked at
\begin{equation}
\tau_{max}=\tau_i-\tau_j-\tau_k+\tau_l
\label{EQtau}.
\end{equation}
With enough visibility pairs, we can produce a set of coupled linear system equations like Equation \ref{EQtau} so that we can solve for all $\tau$ simultaneously.\footnote{This system of equations has a degenerate additive offset (an overall phase) that cannot be solved without an additional constraint. This is equivalent to increasing the length of the cables connecting each tile by the same length. Since this term drops out of any difference $\tau_i-\tau_j$, is is not physically meaningful and can be fixed arbitrarily (e.g.\ by demanding that all delays average to 0).}
Multiplying each $v_{ij}$ by $e^{-2{\pi}i\nu(\tau_i-\tau_j)}$ flattens the phase across the band. This gives us a reasonably accurate starting point for later calibration that effectively avoids phase wrapping. Since all we require is a reasonable starting point for \texttt{OMNICAL}, it is unnecessary to include all redundant baseline pairs into the calculation. The number of all redundant baseline pairs is large (27032 pairs), while a subset of baseline pairs can be sufficient. We only include baseline type (1001,1005) and (1001,1006) (see Figure \ref{FIGpos}), so that the number of baseline pairs is reduced by a factor of 10 (2970 pairs), which is more computationally efficient.\\

\subsubsection{Degeneracy Projection}
\label{DPSec}

Since redundant calibration does not rely on any information from the sky, there are 4 intrinsic degeneracy parameters per frequency per polarization that \texttt{OMNICAL} cannot constrain: one overall amplitude, which depends on the sky flux density; one absolute phase, which depends on the absolute timing of incoming plane waves; and two rephasing parameters, which corresponds to the tip and tilt of the array, or equivalently, the location of the phase center on the sky \citep{Liu2010,Zheng2014,dillon2017redcal}.  \texttt{OMNICAL} can only be performed in the redundant subset of the MWA Phase II array (71 tiles), and without the degeneracy parameters determined, \texttt{OMNICAL} alone cannot provide an absolute calibration. To perform absolute calibration after \texttt{OMNICAL}, we use the FHD calibration solutions as references to constrain the degeneracy parameters. Since FHD calibration is based on a sky model, we take the knowledge of the sky flux density and sky center of FHD as a fiducial guess. We then look for the best fit 4 degeneracy parameters per frequency per polarization for the whole array for the \texttt{OMNICAL} solutions which makes them comparable to FHD results. This fitting process is defined as degeneracy projection.

The fitting for the amplitude parameter is straightforward: for \texttt{OMNICAL}, multiplying all $g_i$ by an arbitrary positive constant, and simultaneously dividing $y_{ij}$ by the square of that constant does not change the amplitude of $g^*_ig_jy_{ij}$. We correct the amplitude degeneracy parameter by multiplying each \texttt{OMNICAL} gain by $e^{\delta}$, where

\begin{equation}
\label{EQampdp}
\delta = \frac{1}{N_{tiles}}\left(\sum_i\eta_i^{FHD} - \sum_i\eta_i^{OMNICAL}\right)
\end{equation}

To illustrate phase degeneracies, we evaluate Equation \ref{EQphsd}, which is the phase part of Equation \ref{EQassumption}: 
\begin{equation}
\gamma_{ij}=\phi_i-\phi_j+\theta_{ij},
\label{EQphsd}
\end{equation}
where $\gamma_{ij}\equiv arg(v_{ij})$, $\theta_{ij}\equiv arg(y_{ij})$. We can add a linear field $\vec{\Phi}\cdot{\vec{r}_i}+\psi$ to $\phi_i$, and simultaneously subtract $\vec{\Phi}\cdot{(\vec{r}_i-\vec{r}_j)}$ from $\theta_{ij}$, to get a new set of solutions as defined in Equation \ref{EQtrans}:
\begin{equation}
\begin{cases}
\phi_i^\prime=\phi_i+\vec{\Phi}\cdot{\vec{r}_i}+\psi\\
\theta_{ij}^\prime=\theta_{ij}-\vec{\Phi}\cdot{(\vec{r}_i-\vec{r}_j)}\\
\end{cases}
\label{EQtrans}
\end{equation}
Under this transformation, $\gamma_{ij}$ in Equation \ref{EQphsd} is invariant, as Equation \ref{EQdegen} shows \citep{Zheng2016}. 

\begin{equation}
\begin{split}
\phi_i^\prime-\phi_j^\prime+\theta_{ij}^\prime=&\ (\phi_i+\vec{\Phi}\cdot{\vec{r}_i}+\psi)-(\phi_j+\vec{\Phi}\cdot{\vec{r}_j}+\psi)\\
&+(\theta_{ij}-\vec{\Phi}\cdot{(\vec{r}_i-\vec{r}_j)})\\
=&\ \phi_i-\phi_j+\theta_{ij}
\end{split}
\label{EQdegen}
\end{equation}
Here $\vec{r}_i$ is the ideal position of tile $i$, i.e., tile positions with perfect redundancy. 
We assume all tiles are coplanar, $\vec{r}_i$ is a 2D vector, thus $\vec{\Phi}$ is 2D.
The absolute phase parameter is given by $\psi$, and the two rephasing parameters are given by the 2D vector $\vec{\Phi}$. 

We define $\Delta{\Psi_i}\ {\equiv}\ \arg(g_{i}^{FHD}/g_{i}^{OMNICAL})$. Equation \ref{EQdplane} shows the relation between calibration solutions and phase degenerate parameters.

\begin{equation}
\Delta{\Psi}_i=\Phi_xx_i+\Phi_yy_i+\psi
\label{EQdplane}
\end{equation}
where $(x_i,y_i)=\vec{r}_i$. Equation \ref{EQdplane} is a function of a plane. The basic idea of solving for ($\Phi_x$, $\Phi_y$, $\psi$) is to fit a plane in ($x_i$, $y_i$, $\Delta{\Psi}_i$) space. This is the process of phase degeneracy projection. The fitting details are described in Appendix \ref{DPASec}. 

\section{Observations of ORBComm}

\label{orbcommSec}
\begin{figure}
 \centering
\includegraphics[width=\linewidth]{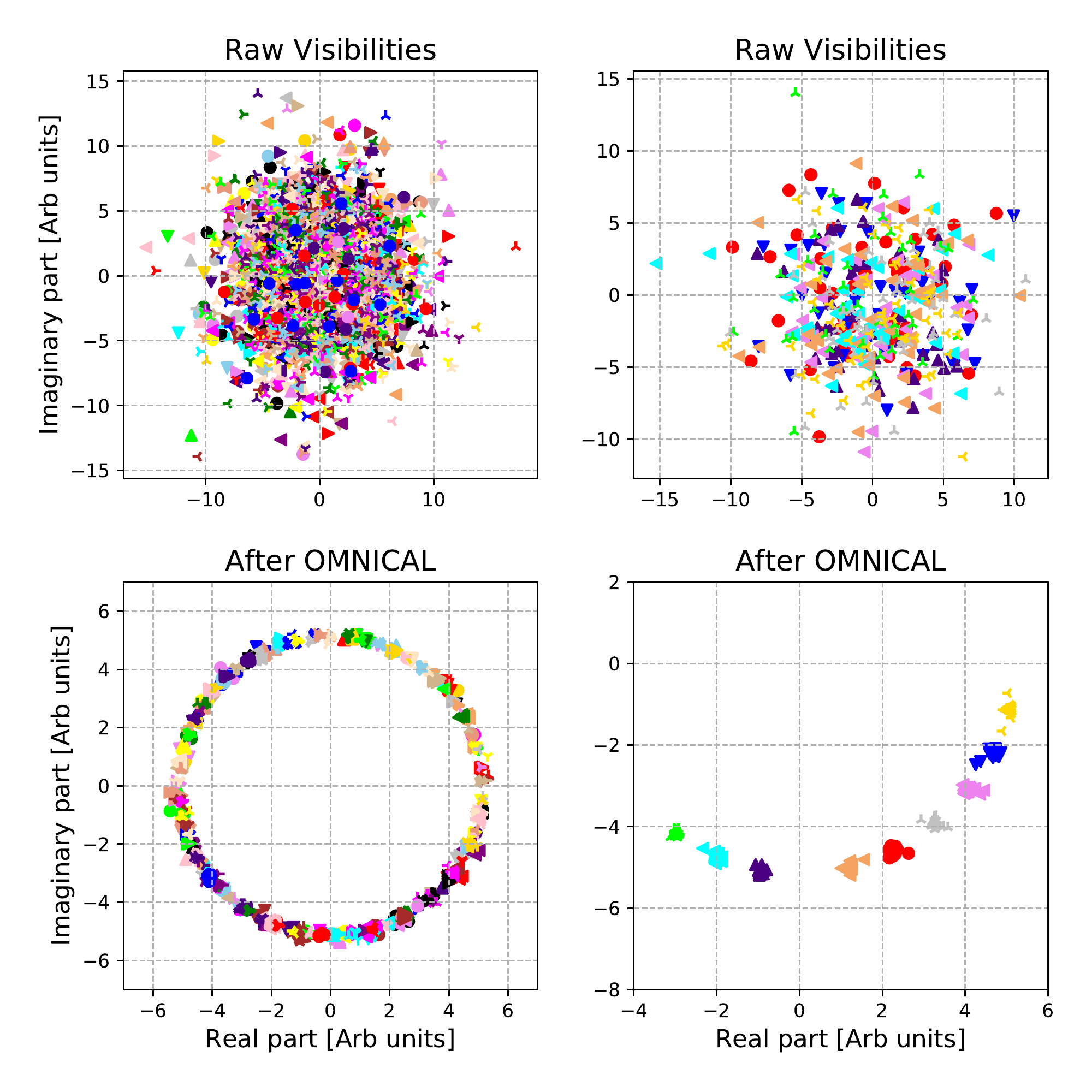}
 \caption{Complex visibilities plots of ORBComm observation at 137.1 MHz, 4 seconds of data. Each unique combination of color and symbol represents visibility measurements from a unique baseline type. Upper left: Raw visibilities from all redundant baseline groups; Bottom left: Calibrated visibilities from all redundant baseline groups; Upper right: Raw visibilities from 9 baseline types; Bottom right: Calibrated visibilities from 9 baseline types. The units are arbitrary because no absolute calibration is performed.}
 \label{FIGORBComm}
 \end{figure}

As a first test of \texttt{OMNICAL} on MWA Phase II data, we investigate observations at 137.1 MHz where the ORBComm satellite system transmits \citep{neben2015measuring,neben2016hydrogen,ORBCommLine} because this data set has extremely high signal-to-noise.
Since the MWA has a wide field of view, it is difficult to point to a patch of sky with a flux density dominated by one bright point source. However, an ORBComm satellite provides a good opportunity to observe a `point source' because its signal is orders of magnitude brighter than any other sources in the sky at 137.1 MHz. The near-infinite signal-to-noise measurements on ORBComm are an excellent opportunity to quantify the uncertainties in the redundant calibration procedure \citep{Zheng2014}. 

Figure \ref{FIGORBComm} shows the \texttt{OMNICAL} results on observations of an ORBComm satellite with the MWA Phase II hexagons on Sept 21, 2016. Each unique combination of color and symbol represents visibilities measured by a redundant baseline group. 
The upper left plot shows the complex visibilities from all redundant baseline groups before \texttt{OMNICAL}, and the lower left shows the same set of data after calibration.

The constant amplitude of visibilities in the lower left plot indicates a delta function in the image domain, which agrees with our point source expectation of ORBComm. We pick 9 unique baseline groups as representatives from the left column in Figure \ref{FIGORBComm} and show the uncalibrated (upper right) and calibrated (bottom right) visibilities in the right column. This illustrates that \texttt{OMNICAL} makes visibility measurements from baselines with the same length and orientation cluster together, i.e., \texttt{OMNICAL} is performing as expected: it makes visibilities from physically redundant baselines agree with each other. The level of the standard deviation within each redundant visibility group is 1\% comparing to their magnitudes, which is possibly due to the non-perfectly gridded antenna positions. This quantifies the systematic uncertainty of redundant calibration procedure for MWA PhaseII array, or in other words, this level of disagreement is the best that redundant calibration can achieve.\\

\section{Comparison between FHD and \texttt{OMNICAL}}
\label{comparisonSec}
In this section, we will take MWA Phase II observations targeting the EoR0 field as an example to show the comparison between FHD sky calibration and redundant calibration. 
All calibrations are performed per data file (every 112 seconds), and the gains are assumed to be time independent within a data file. In FHD calibration, the sky model is a point source catalog specifically developed for EoR0 field \citep{Carroll2016}. All time steps (2 second integrations) are fed into the linear least-square solver which minimizes the difference between data and model and returns one set of time-independent calibration solutions per file.

\begin{figure*}
 \centering
 \includegraphics[width=\linewidth]{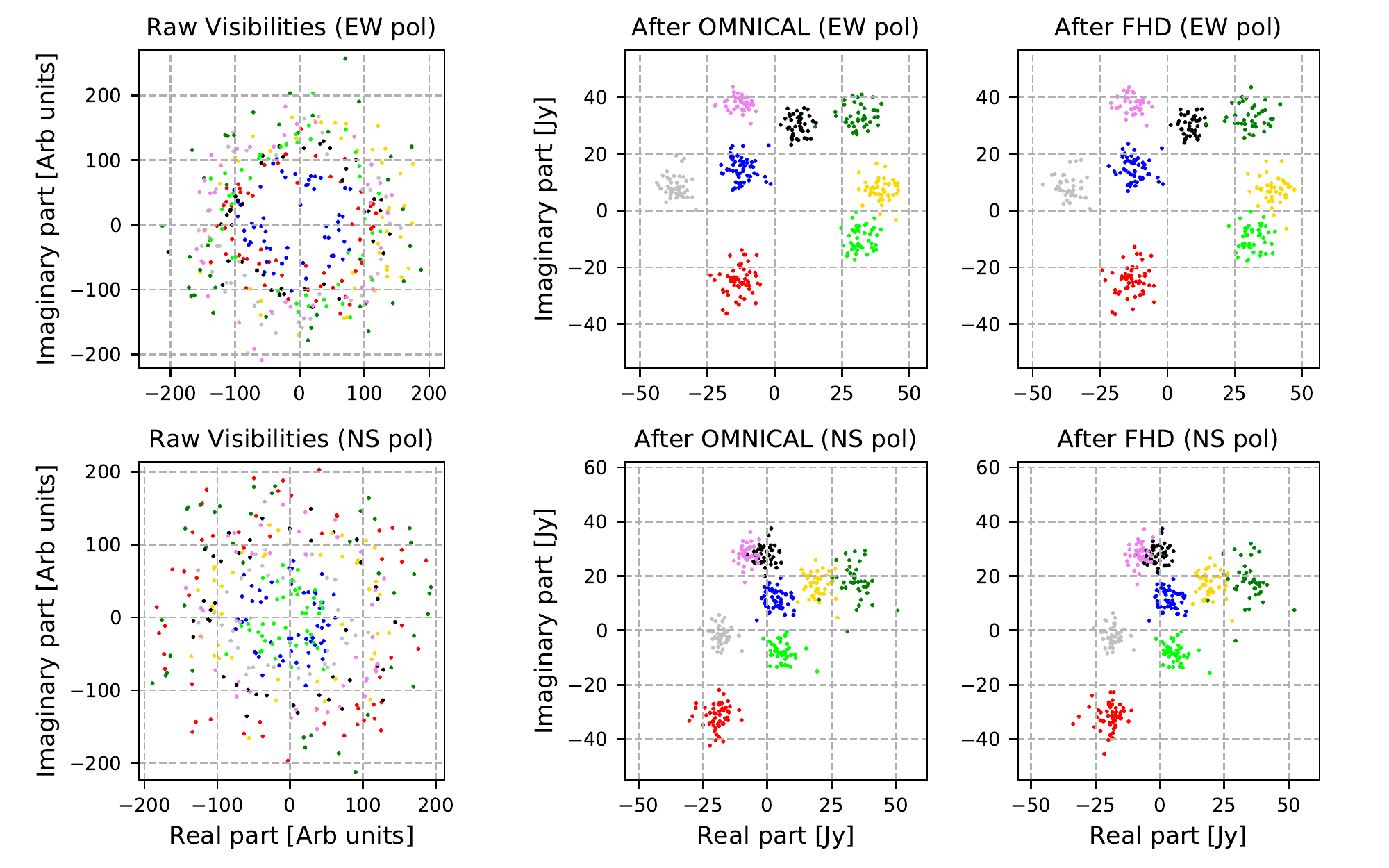}
 \caption{112 seconds averaged complex visibilities plots of EoR0 data at 191 MHz from 8 redundant baseline groups. Each color represents one redundant baseline group. Left column: raw visibilities, with arbitrary units; Middle column: visibilities after \texttt{OMNICAL}, with degeneracy parameters projected (units: Jy); Right column: visibilities after FHD sky calibration (units: Jy). Top row: East-West polarization; Bottom row: North-South polarization.}
 \label{FIGcluster}
\end{figure*}

In \texttt{OMNICAL}, we average the data along the time axis of each data file, i.e., average every two minutes of data before calibrating for two technical purposes: increasing SNR for better redundant calibration performance \citep{Liu2010} and excluding sparse flagged data samples without dramatically increasing computational expense.\footnote{In \texttt{OMNICAL}, explicitly excluding flagged baseline samples per time and frequency requires generating distinct linear equations per time and frequency instead of per data file, which is computationally infeasible.}

We also exclude the baseline type (1001, 1002) (the index refers to Figure \ref{FIGpos}), which is the 14 m east-west baseline type, because we have seen significant systematics from that baseline group. The visibility variances in this group are about 6 times larger than other redundant baseline groups. This could be due to strong cross-talk between tiles, or because the Galactic plane aligns with these baselines (c.f. \citealt{Thyagarajan2015}), enhancing the effect of tile-to-tile beam variations across the array \citep{noorishad2012redundancy}. The reason is still unclear, but it is a topic to be investigated in future work.

\subsection{Visibility Clustering}

The redundant baselines should measure the same Fourier mode of the sky regardless of the calibration procedures involved. Evaluating how visibilities measured by redundant baselines agree with each other (visibility clustering) is an approach to evaluate calibration methods. Figure \ref{FIGcluster} shows 112 second averaged complex visibilities at 191 MHz observed on Nov 21, 2016. We plot the visibilities for 8 types of baselines with lengths below 20 wavelengths, which are of most importance for EoR sensitivity, at 180 MHz. Visual inspection shows substantial agreement between the two methods.  Quantitatively, visibilities after \texttt{OMNICAL} (middle column) are in better agreement than FHD (right column) (about 6\% to 30\% reduction in the standard deviation of a cluster). One explanation for this effect is that in FHD calibration, baselines shorter than 50 wavelengths at 180 MHz are omitted (due to the difficulties in modeling diffuse emission; \citealt{patil2016systematic}).  Thus, short baselines (like those plotted here) have less weight in FHD calibration; \texttt{OMNICAL} uses the information of these short baselines. \texttt{OMNICAL} also explicitly minimizes the variance within redundant visibilities, thus it should lead to better visibility clustering than alternative methods. Although this metric does not necessarily indicate a better calibration, it shows that it is possible to put more weight on the most EoR-sensitive baselines, instead of calibrating with only long baselines with low EoR sensitivity as is currently required for sky calibration.

\begin{figure*}
 \centering
 \includegraphics[width=\linewidth]{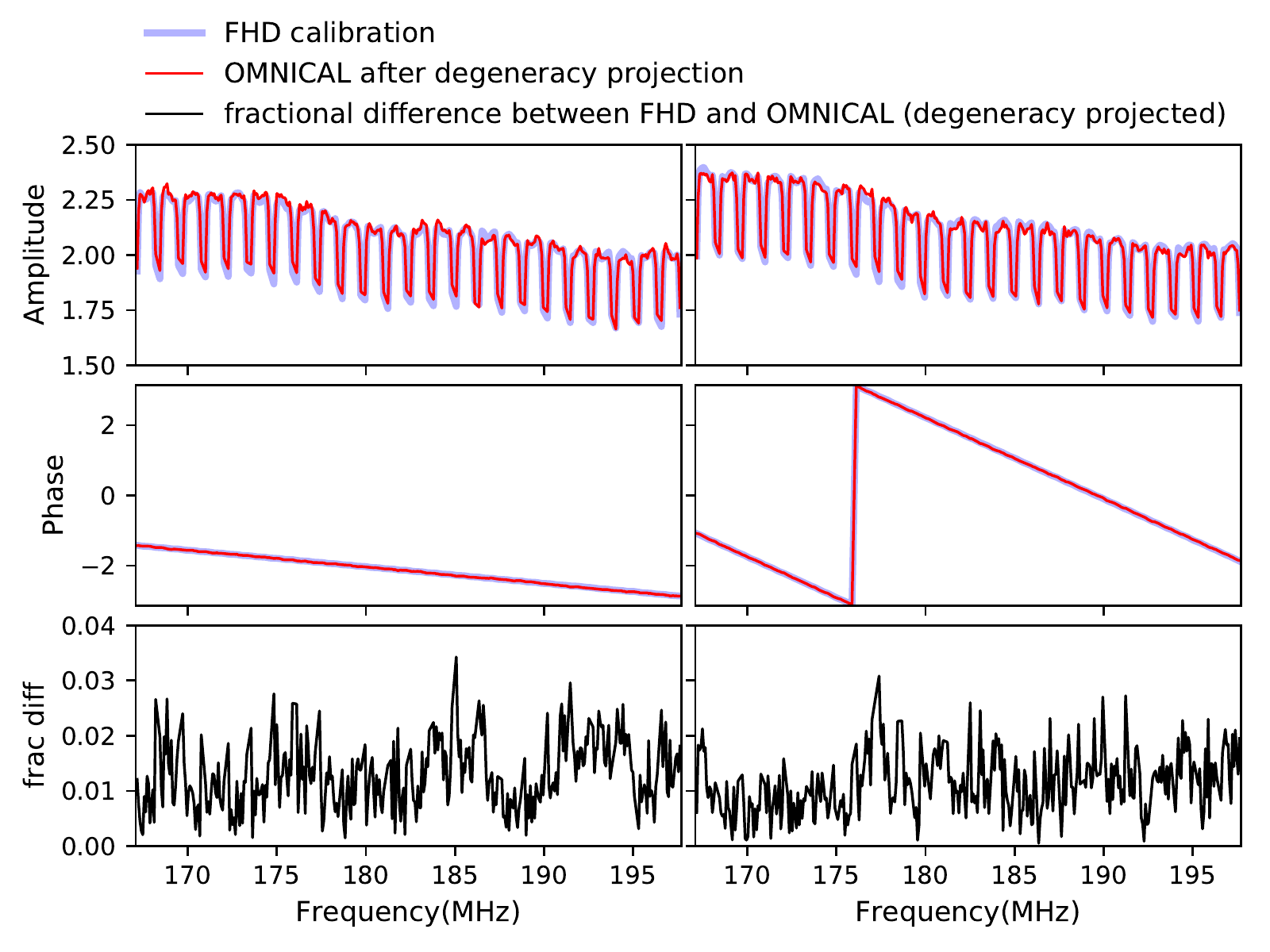}
 \caption{30 minutes averaged gain solutions of tile 1024 (left column) and tile 1064 (right column) from zenith pointing, east-west polarization. Upper: Gain amplitude; Middle: Gain phase. Lower: fractional difference between FHD solution and \texttt{OMNICAL} solutions with degeneracy projected. Blue: FHD solutions; Red: \texttt{OMNICAL} solutions after projecting degeneracy. The fractional difference in the lower plots is calculated by dividing the amplitude of the complex difference between the two by the amplitude of FHD solutions.}
 \label{FIGproj}
\end{figure*}

\subsection{Direct Comparison}

Figure \ref{FIGproj} shows a direct comparison between FHD solutions and \texttt{OMNICAL} solutions after degeneracy projection.  The data are from a 30 minute zenith pointing on the EoR0 field, and calibration solutions have been averaged over the entire pointing.  Figure \ref{FIGproj} shows solutions for tile 1024 and tile 1064 in gain amplitudes (top) and phases (middle) over frequency, as well as fractional difference between solutions from these two approaches (bottom). 

The first conclusion is the bandpass structures from both approaches show consistent results at a level of 98\%. 
However, the 2\% level of difference between the two is not negligible as far as the EoR signal is concerned. For each 1.28 MHz sub-band, the frequency channels near the band edges appear to show relatively larger differences. The differences in solutions can come about not only because they are derived with different algorithms using different assumptions, but also because they use different subsets of the data to perform the calibration, i.e., FHD uses data from long baselines, while \texttt{OMNICAL} uses data from redundant baselines. We will investigate the effects of this level of difference on EoR PS measurements in Section \ref{combiningSec}.

\section{Combining FHD and \texttt{OMNICAL}}
\label{combiningSec}
FHD performs well on calibrating EoR0 data with a well developed point source catalog (e.g. \citep{Beardsley2016,barrythesis}, but it also has shortcomings, including errors introduced by an incomplete sky model \citep{Barry2016} and a loss of sensitivity from excluding short baselines due to difficulty in modeling diffuse sources \citep{patil2016systematic,Sullivan2012,bowman2009foreground}. 
\texttt{OMNICAL} is free of sky model error and is able to calibrate short baselines (although, as noted, we exclude the shortest 14 m east-west baselines from redundant calibration because they exhibit significantly larger scatter than other redundant baseline types), but it cannot solve for the degenerate parameters, and it can only calibrate a subset of the array. In addition, \texttt{OMNICAL} has the potential for error introduced by tile position inaccuracies and beam variation from tile to tile. 

Their respective advantages and disadvantages, however, suggest that \texttt{OMNICAL} and FHD can be mutually complementary. We can possibly use the algorithms to mitigate both sky model and non-redundancy errors. These two methods also allow us to make use of more baselines for calibration, since FHD excludes short baselines and \texttt{OMNICAL} only can calibrate antennas in the redundant subset of the array.

With bad tiles excluded (tile 45 and tile 1037 are not operational in our data set), there are 71 hexagon tiles and 55 non-hexagon tiles. In FHD calibration, if we only calibrate baselines longer than 50 wavelengths at 180 MHz, the number of baselines we use is 5653. For the combined calibration, there are 2477 baselines involved in redundant calibration,  1235 of them are shorter than 50 wavelengths, thus 6888 baselines can be used in calibration.

In this section, we propose two strategies to combine FHD with \texttt{OMNICAL}. As our metric for evaluating different approaches, we use the two-dimensional $(k_\perp,\ k_\parallel)$ power spectrum common to 21 cm EoR analyses.\footnote{Our power spectrum estimator (discussed below) uses all baselines, so it is necessary to combine both FHD and \texttt{OMNICAL} to get calibration solutions for both the hex and non-hex tiles.  Hence, we do not use the PS metric to compare the independent solutions from FHD and \texttt{OMNICAL} in the previous section.} 
A schematic 2D PS is shown as the upper left plot in Figure \ref{FIGps}. The power in the lower red region in $k_\parallel$ for all $k_\perp$ is dominated by the intrinsically spectrally smooth foregrounds. The instrument chromaticity mixes foreground modes up to higher $k_\parallel$, forming into a `foreground wedge'. The limit of the wedge depends on how far the sources are from the center of the field of view and increases on longer baselines (larger $k_\perp$). The solid line and dashed line represent the horizon limit and the primary field of view limit, respectively. The remaining `EoR window' is foreground free and expected to contain a wealth of information about the 21 cm signal 
\citep{Barry2016,Datta2010,Morales2012,Vedantham2012,Parsons2012, Trott2012,Hazelton2013,Thyagarajan2013,Pober2013,Liu2014}. In the `EoR window', any observed excess of power is a contaminant, as the EoR signal is buried deep in the noise. Our metric of evaluating calibration techniques is to quantify their performances of mitigating power contamination in the `EoR window'. Not only is this metric the quantity of interest (a major goal of MWA Phase II is to measure the PS of the EoR), it also highlights subtle differences between the calibration schemes due to its inherent sensitivity to spectral structure which can corrupt EoR measurements.

To create our PS, we use the software package Error Propagated Power Spectrum with Interleaved Observed Noise (\texttt{$\epsilon$ppsilon}\footnote{https://github.com/EoRImaging/eppsilon} which calculates the PS using image cubes as input with errors propagated through the full analysis; \citealt{Jacobs2016}).

\subsection{Strategies}
\label{sec:combine_strategies}
We propose two simple strategies to combine \texttt{OMNICAL} with FHD by running them sequentially: ``\texttt{OMNICAL} first, FHD second" and ``FHD first, \texttt{OMNICAL} second." To simplify, we name \texttt{OMNICAL} first, FHD second as OFcal, and FHD first, \texttt{OMNICAL} second as FOcal. Since \texttt{OMNICAL} can only calibrate the subset of the array, no matter what strategy we propose, these hybrid approaches only change the calibration on hexagon tiles; the calibration of non-hexagon tiles remains the same as FHD calibration results.

\begin{figure}
\centering
\includegraphics[width=\linewidth]{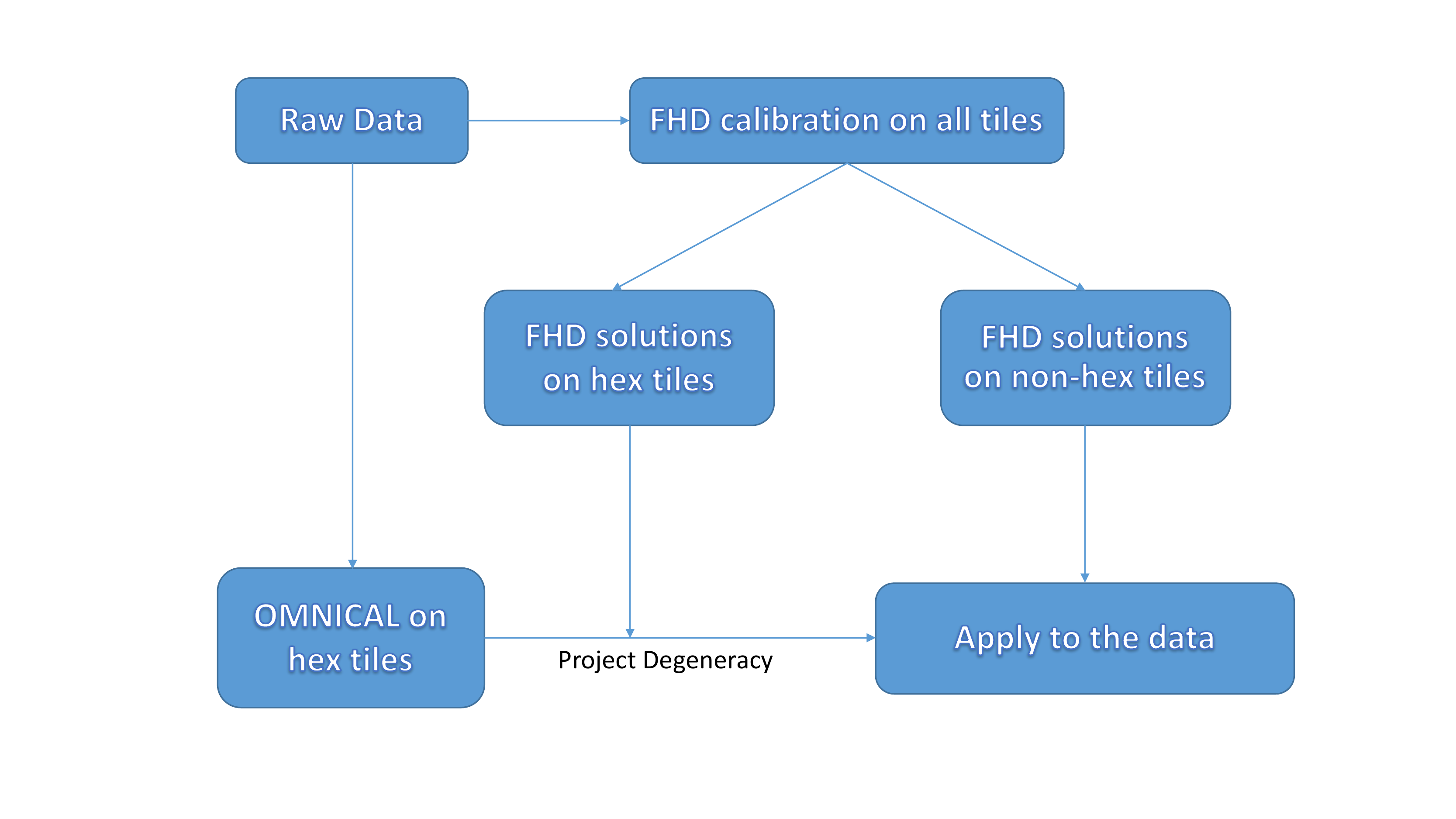}
\caption{Flow diagram showing the procedure of OFcal.}
\label{FIGOF}
\end{figure}

\begin{figure*}
\centering
\includegraphics[width=\linewidth]{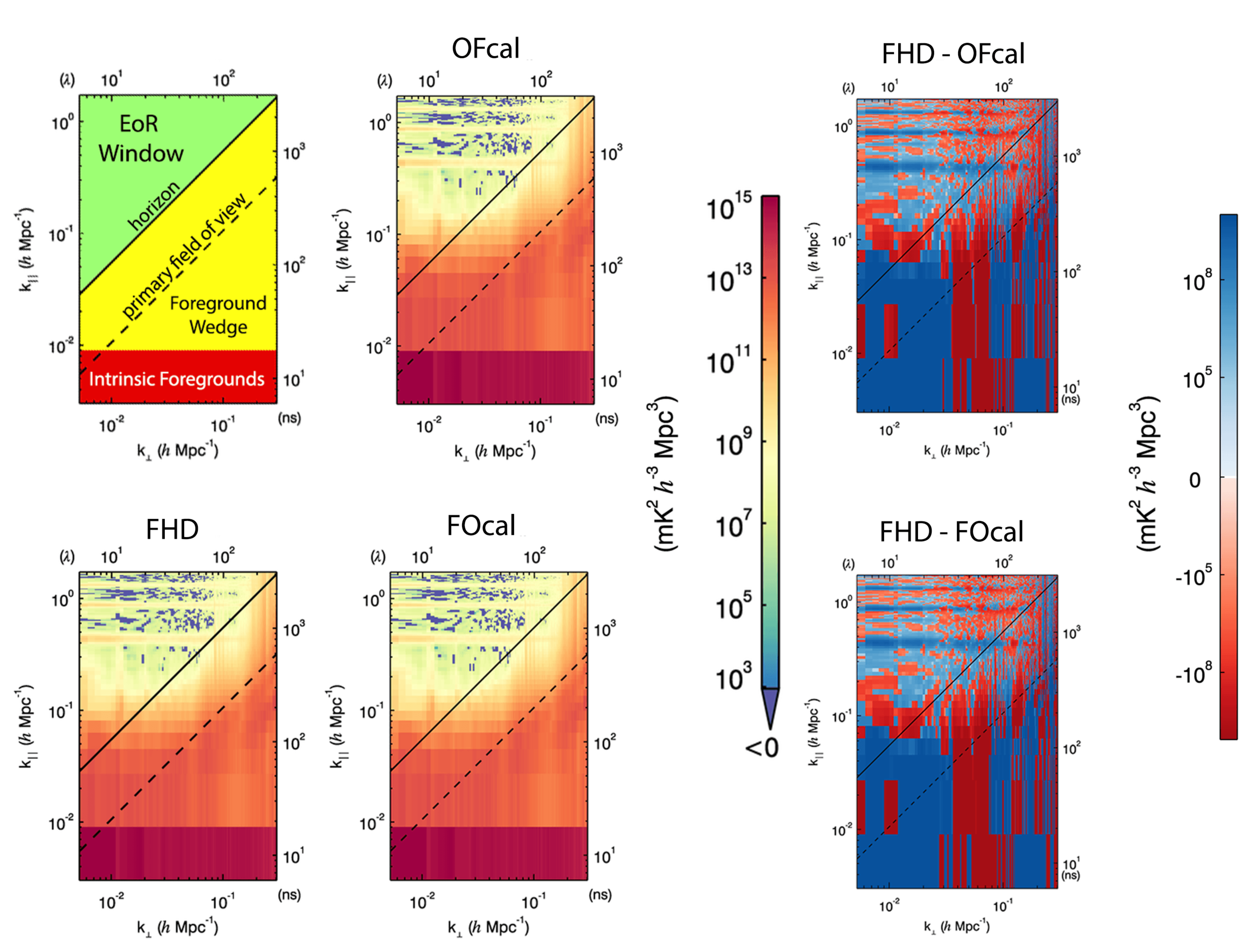}
\caption{Upper left: schematic plot of 2D cylindrical power spectrum. Low $k_\parallel$ modes are dominated by intrinsic foregrounds and the chromaticity of the interferometric instrument smears foregrounds contamination up to high $k_\parallel$, leaving an `EoR window' which is foreground free. Lower left: PS after FHD calibration. Upper middle: PS after OFcal. Lower middle: PS after FOcal. Upper right: difference PS of FHD minus OFcal. Lower right: difference PS of FHD minus FOcal.  See text for details on the calibration methods.}
\label{FIGps}
\end{figure*}

\subsubsection{OFcal}
The OFcal approach is illustrated in the diagram shown Figure \ref{FIGOF}. The calibration procedure is as follows:
\begin{enumerate}
\item Run \texttt{OMNICAL} on raw visibilities measured by baselines within hexagon tiles;
\item Perform FHD calibration on raw visibilities measured by all baselines longer than 50 wavelengths at 180 MHz;
\item Average \texttt{OMNICAL} solutions for each pointing (30 minute time average); 
\item Project degeneracy parameters of \texttt{OMNICAL} solutions to FHD solutions;
\item Apply degeneracy-projected, time-averaged \texttt{OMNICAL} solutions to tiles within the hexagons and apply FHD solutions to all other tiles.
\end{enumerate}
When averaging calibration solutions from a single pointing, we first make sure this set of solutions have same degeneracy parameters. We do this by picking one data file solutions as target, and projecting degeneracy of solutions from other data files to this target, then average.

\subsubsection{FOcal}
The description of FOcal is simpler:

\begin{enumerate}
\item Perform FHD calibration on raw visibilities measured by all baselines;
\item Apply FHD solutions to the raw data;
\item Run \texttt{OMNICAL} on FHD calibrated visibilities (from baselines within the hexagons);
\item Average \texttt{OMNICAL} solutions for each pointing (30 minute time average);
\item Project degeneracy parameters of \texttt{OMNICAL} solutions to default values of 0;
\item Apply time-averaged \texttt{OMNICAL} solutions to FHD calibrated data.
\end{enumerate}

Since the data before \texttt{OMNICAL} is already calibrated by FHD, the degeneracy is removed in a different way. By forcing the average of $\eta$'s of all tiles to be 0 (similar to Equation \ref{EQampdp}, but with FHD terms excluded), the flux density scale set by FHD does not change, and by making the linear field $\Phi$ have zero slope and setting the average phase to be 0, the sky center does not change. More details are described in Appendix \ref{DPASec}.

The basic difference between OF and FOcal is that each individual baseline has different weights in these two cases. When constructing $\chi^2$ for \texttt{lincal}, with each individual term to be $|v_{ij}-g_ig_j^*y_{ij}|^2$, we will see baselines with larger $|g|$ having larger noise level. In OFcal, to avoid any bias, we weight each $v_{ij}$ by the reciprocal of product of the square root of autocorrelations of tile $i$ and tile $j$, which effectively cancels out the gain amplitude differences. In FOcal, since the amplitude calibration of FHD is already applied before \texttt{OMNICAL}, we do not apply this weighting. In the noiseless case, we expect both approaches to yield the same result, but in the presence of noise the best weighting for these methods is an open question.

\subsection{Results}
\label{sec:combine_results}

A PS comparison of OFcal and FOcal with the FHD-only calibration for the North-South polarization data is shown in Figure \ref{FIGps}. The PS of the data with only FHD calibration applied is shown in the lower left panel.  The middle column shows the PS of the data with the two new calibration schemes applied (OFcal on top, and FOcal on the bottom). 

\begin{figure}
\centering
\includegraphics[width=\linewidth]{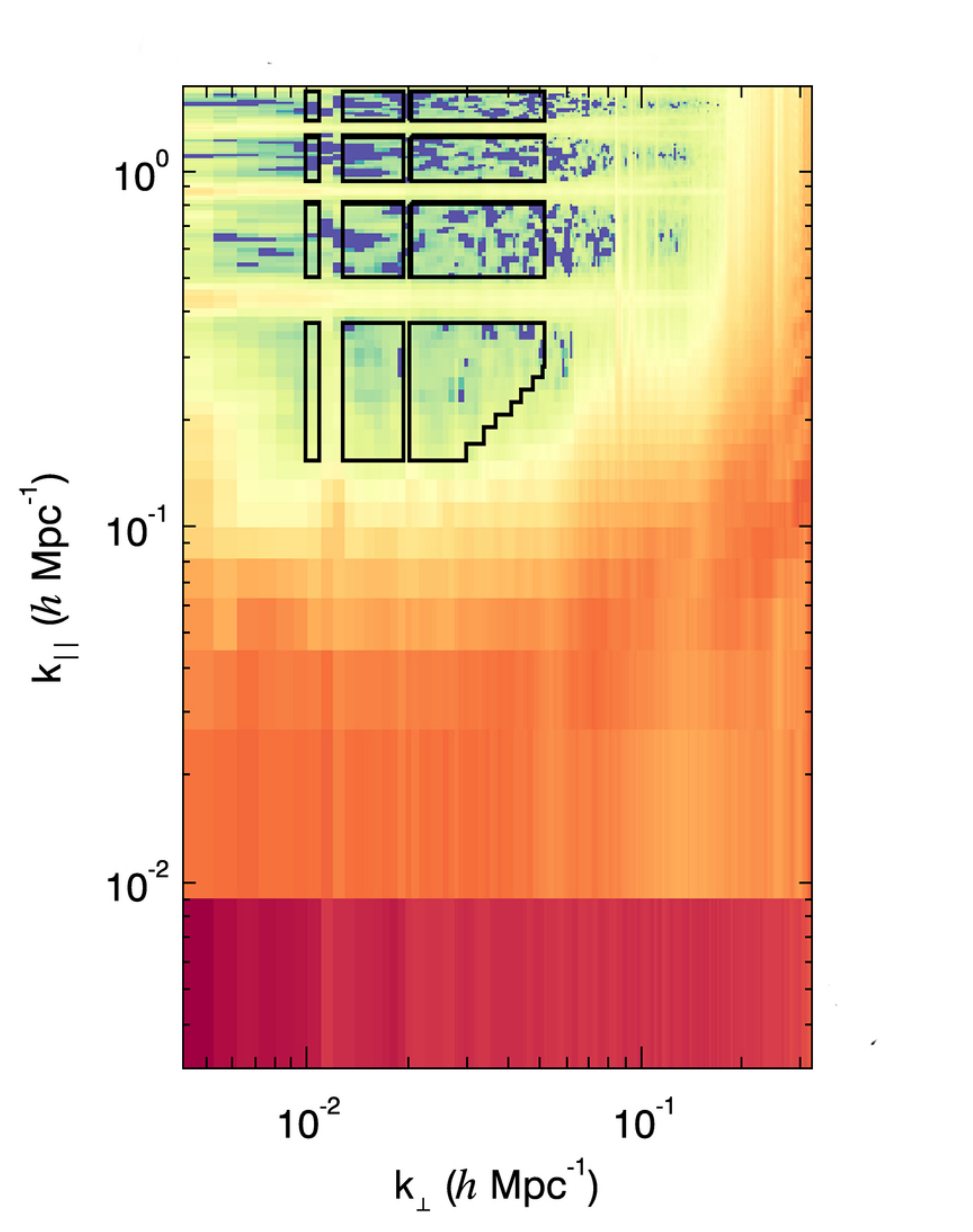}
\caption{The 2D power spectrum using calibration from FHD only (Figure \ref{FIGps}, lower left) with contours to highlight modes that will be used for 1D power in $k_\parallel$ in Figure \ref{FIG1d}}
\label{FIGmodes}
\end{figure}

\begin{figure*}
\centering
\includegraphics[width=\linewidth]{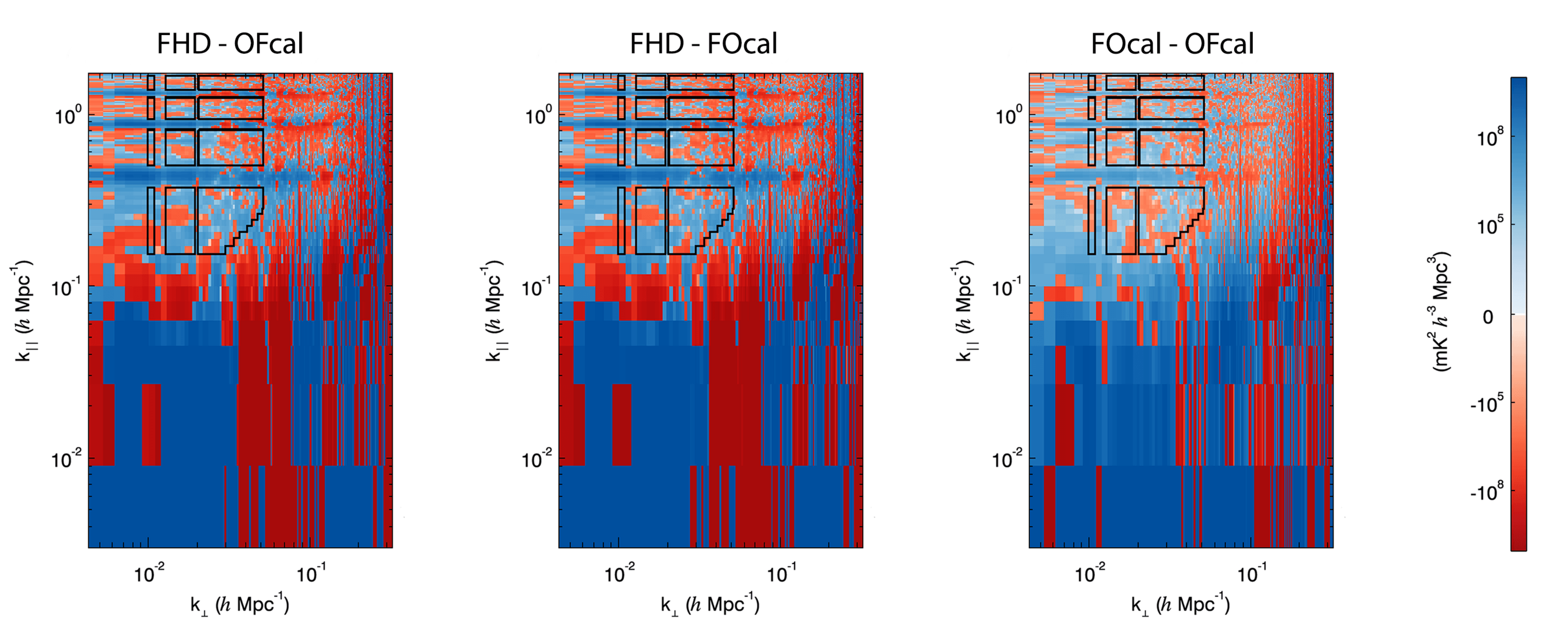}
\caption{ps difference plots of FHD minus OFcal (left; Figure \ref{FIGps}, upper right), FHD minus FOcal (middle; Figure \ref{FIGps}, lower right) and FOcal minus OFcal (right) with the same contours as Figure \ref{FIGmodes}.}
\label{FIGcontour}
\end{figure*}

The three PS from FHD, OFcal and FOcal have common features and are nearly indistinguishable. The horizontal streaks of excess power shown in PS plots are harmonic modes due to flagged channels between every 1.28 MHz sub-band. The vertical streak at $\sim 12$ wavelengths is due to sparse sampling in $k_\perp$ space, or in other words, we do not have baselines sampling those modes. 

To illustrate the difference between OFcal (or FOcal) and FHD, we make difference PS plots shown in the right hand side of Figure \ref{FIGps}. The difference plots (right column) are obtained by subtracting (in 3D $k$ space) the PS of data with OFcal applied (upper right) and FOcal applied (lower right) from that of the FHD-only calibrated data. In PS difference plots, red indicates an excess of power in the OFcal (or FOcal) strategy and blue indicates a reduction of power when compared with FHD. From the difference plot, we can conclude that both OFcal and FOcal show lower power at sub-band harmonic modes than FHD. We expect this improvement at sub-band harmonic modes because the channels near sub-band gaps seem to show the most tile to tile variation. \texttt{OMNICAL} is capable of capturing this variation, while FHD only fits a smooth polynomial functions in frequency (after dividing by a cable-averaged bandpass) to capture tile to tile variation (see Section \ref{fhdcalSec}).  Because these variations appear on the sub-band scale of 1.28 MHz, FHD cannot calibrate them out as well as \texttt{OMNICAL}. 

\begin{figure*}
\centering
\includegraphics[width=\linewidth]{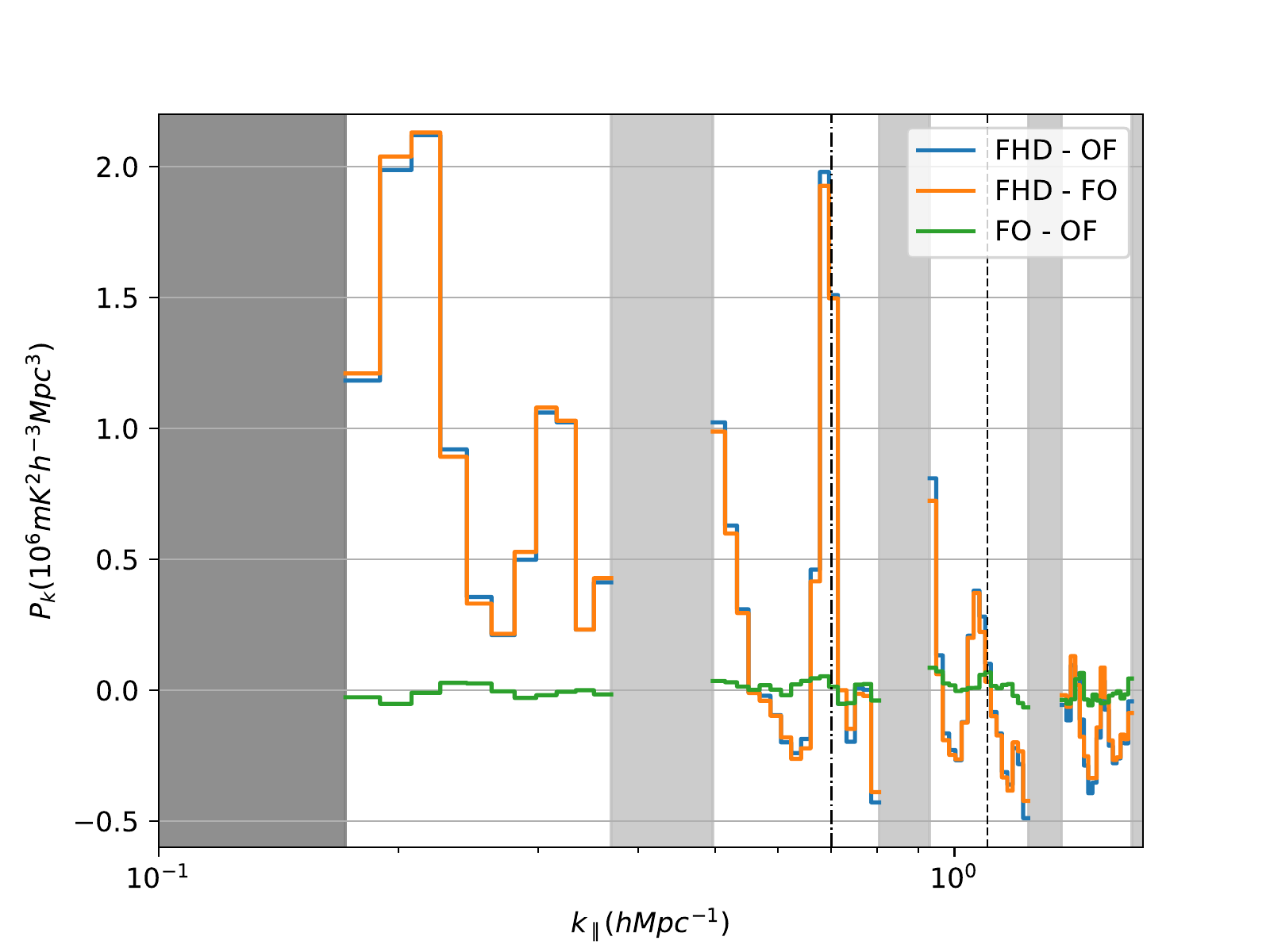}
\caption{1D difference power spectrum versus $k_\parallel$ made from the subset of modes illustrated in Figure \ref{FIGcontour}. Blue: FHD only minus OFcal; Orange: FHD only minus FOcal; Green: FOcal minus OFcal. The dark gray shaded region indicates low $k$ modes which are foreground contaminated; the light gray shaded regions are sub-band harmonic modes we cut out. The vertical dash-dot line and dashed lines highlight the 150 m and 230 m cable reflection modes, respectively.}
\label{FIG1d}
\end{figure*}

To further investigate the PS differences in the EoR window, we pick regions of $k$ space which are free from foregrounds and sub-band contaminations in the 3D power spectrum.\footnote{Similar selections were used in \cite{Beardsley2016}, although we exclude select values of $k_\perp$ where the Phase II baseline sampling is poorer than in Phase I.} We illustrate these cuts in Figure \ref{FIGmodes}, where the contamination we are excluding is evident.  To more clearly demonstrate improvements from the combined calibration technique, we apply this $k$ space cut to the power spectrum differences shown in Figure \ref{FIGcontour} and average in $k_\perp$ to make a 1D power difference versus $k_\parallel$, which we show in Figure \ref{FIG1d}. We have excluded low $k_\parallel$ modes which are foreground contaminated (dark gray), as well as sub-band harmonic modes (light gray).  Figure \ref{FIG1d} shows that OF$/$FOcal both show less contamination than FHD in general (i.e. the differences in Figure \ref{FIG1d} are mostly positive). Both hybrid approaches show better performance at 150 m reflection modes (the expected $k_{\parallel}$ value for a 150 m cable reflection at redshift 7 is marked by the vertical dot-dashed line; \citep{ewall2016first}). We can also see some improvements near the 230 m cable reflection mode marked by the vertical dashed line.

We also see that both OFcal and FOcal have strikingly similar differences with FHD (i.e. the two difference PS in the right-hand side of Figure \ref{FIGps} are very similar), although they are not identical. A difference PS plot of FOcal minus OFcal is shown in Figure \ref{FIGcontour} on the right, and the 1D version of this PS difference is shown in Figure \ref{FIG1d} (green). We conclude the differences between OFcal and FOcal are centered around zero and much less significant overall.

We note that in all cases, these differences are \emph{below} the thermal noise level in our measurements.  However, in creating these difference power spectra, we are subtracting the same data --- with the exact same realization of the noise --- only with different calibrations applied.  If the calibrations were the same, the differences would be identically zero.  Since the goal of these experiments is to detect the 21\,cm signal from the EoR, the typical amplitude of the EoR signal --- approximately $10^6\ \mathrm{mK}^2h^{-3}\mathrm{Mpc}^3$ at $k \sim 0.1\ h{\rm Mpc}^{-1}$ \citep{furlanetto2006,mesinger2011} --- provides a rough scale for assessing the significance of our improvements.  Using redundant calibration in addition to FHD (either through FOcal or OFcal) removes foreground contamination at or above the level of the EoR signal.  The differences between FOcal and OFcal are much smaller and are thus unlikely to be significant for EoR experiments.

\section{Discussion}
\label{discussionSec}
Section \ref{sec:combine_results} has shown that our hybrid approaches (OF$/$FOcal) can improve the power spectrum in the EoR window. The intuition for this improvement is that tiles in the two hexagons were calibrated based on redundant baseline assumption, thus all non-degenerate parameters are then free from the sky model error described in \cite{Barry2016}. The 4 degenerate parameters per frequency channel per time per polarization still are sky model dependent, but overall, we expect to have mitigated the error introduced by imperfect sky model. Additionally, FHD only uses long baselines for calibration, which could potentially overfit gain parameters to noises in long baseline data; however, we are more interested in short baseline data in EoR observations, and we expect OF$/$FOcal to mitigate this effect by including short baselines in calibration. Although we are ignoring crosstalk and ionospheric effects in this work, all calibration methods work on the same data and therefore have the same challenges and that nonetheless a noise power reduction was achieved by our hybrid approach.

So far we have not considered systematic errors that can affect \texttt{OMNICAL}. Redundant calibration is based on two assumptions: that redundant baselines have the exact same length and orientation and that all tile beams are identical. These two assumptions are not exactly true in practice. According to MWA Phase II baseline coordinates, position deviations from perfect redundancy is relatively small (at a level of 5 cm). We have performed noiseless foreground simulations to study the effect of systematic errors introduced through redundant calibration using the imperfect tile positions of Phase II. As we see in the simulated data, the so called `redundant' visibilities are not identical, but we assume they are when we do calibration. We found the errors introduced to the power spectrum by the `wrong' redundancy assumption are unbiased as well as below the typical EoR level by 2 orders of magnitude. Beam variation can also be significant. It is also a possible cause for large systematic disagreement for baseline type (1001,1002) (East-West 14-meter-baselines) we saw in this data. In future work, we will explore the error introduced by both effects through detailed calibration simulations, similar to \cite{Barry2016}. 

In our analysis, we performed \texttt{OMNICAL} on each data file after averaging it over time axis (2 minutes), which gives better SNR in calibration and allows us to conveniently avoid flagged samples. However, there is a concern of washing signals out for relatively long baselines. We investigated three averaging scenarios through noiseless foreground simulations using real baseline coordinates where there are no flagged samples: calibrating visibilities each 2 second interval, then directly applying these solutions to the data; averaging calibration solutions derived for each 2 second interval over 2 minutes, then applying the time-averaged solutions to the data; and, most similar to the analysis performed here, averaging 2 minutes of data, then calibrating using the averaged data and applying these to the un-averaged data. By evaluating the power spectrum as we did in Section \ref{sec:combine_results}, we conclude none of the three scenarios show bias relative to others, and the amplitudes of differences among them are 3 orders of magnitudes lower than the typical EoR level. This validates our averaging strategy used in redundant calibration in the real data.

\section{Conclusion}
\label{conclusionSec}

We have explored the application of both sky-based calibration and redundant calibration to data from Phase II of the MWA, and investigated their respective trade-offs and possible complementarity.  Sky calibration is model dependent and a reasonable calibration requires a fairly good model of the radio sky. The sky model, as well as the beam model, cannot be perfect to a certain level. Errors such as wrong source positions, brightness errors, or missing sources, can potentially introduce calibration error to the PS \citep{Barry2016}. In addition, since the sky model in FHD is a point source catalog, and it is difficult to model diffuse sources, the short baselines are omitted \citep{Sullivan2012,patil2016systematic,bowman2009foreground}, which leads to a loss of information of those baselines in calibration. 
Redundant calibration provides an opportunity to remedy these shortcomings: it is sky model independent, thus it is not restricted by baseline length. However, redundant calibration leaves 4 intrinsic degeneracy parameters unsolved. In addition, redundant calibration may also be contaminated by tile position error and beam variation \citep{Liu2010}. Section \ref{sec:combine_results} shows using redundant calibration and sky-based calibration together can alleviate the potential error introduced by assumptions these two approaches made. We aim to make use of the advantages of both calibration approaches and combine them together to improve our calibration. 

In this paper, we have shown the success of \texttt{OMNICAL} on ORBComm observations from MWA Phase II, and compared \texttt{OMNICAL} and FHD on EoR0 data, showing consistent results from these two approaches. This is the first time these two independent methods have been confirmed to agree in real data calibration. We further attempted to combine FHD with \texttt{OMNICAL} in two ways: \texttt{OMNICAL} first, FHD second (OFcal), and FHD first, \texttt{OMNICAL} second (FOcal). By comparing them with FHD in PS scheme, we conclude both OFcal and FOcal show improved behavior in the $k$ modes with the most EoR sensitivity in the power spectrum, especially in modes contaminated by 150\, m and 230\, m cable reflections.\\

This result substantially improves on similar comparisons in the literature.  \cite{noorishad2012redundancy} use redundancy between individual dipole elements within a LOFAR phased-array tile, but the array has little to no redundancy between tiles. \cite{nikolic2017hera} use a point-source model for the Galactic center to calibrate the 19-element, highly redundant HERA commissioning array, but they present no comparisons with redundant calibration methods.  When the complete 350-element HERA is finished, however, it will be a valuable tool for performing studies similar to the one presented here \citep{dillon2016hera}. 

\acknowledgments

WL, JCP, BJH, NB, MFM, and IS would like to acknowledge the support from NSF grants \#1613040, \#1613855, and \#1506024. JSD acknowledges the support of NSF Award \#1701536. APB acknowledges the NSF Astronomy and Astrophysics Postdoctoral Fellowship under award AST-1701440. This scientific work makes use of the Murchison Radio-astronomy Observatory, operated by CSIRO. We acknowledge the Wajarri Yamatji people as the traditional owners of the Observatory site. Support for the operation of the MWA is provided by the Australian Government (NCRIS), under a contract to Curtin University administered by Astronomy Australia Limited. We acknowledge the Pawsey Supercomputing Centre which is supported by the Western Australian and Australian Governments. This research was conducted using computation resources and services at the Center for Computation and Visualization, Brown University. We acknowledge Greg Rowbotham for providing the picture of MWA Phase II tiles.

This work made use of the following open source software: FHD \citep{Sullivan2012}, pyuvdata \citep{JHazelton2017}, \texttt{OMNICAL} \citep{Zheng2014}, hera\_cal (\url{https://github.com/HERA-Team/hera_cal}), $\epsilon$ppsilon \citep{Jacobs2016}.

\appendix
\renewcommand{\appendixname}{Appendix~\Alph{section}}
\section{Degeneracy projection}
\label{DPASec}
Section \ref{DPSec} describes the 4 intrinsic degeneracy parameters per polarization per frequency per time in redundant calibration. In this section, we describe details about how we treat these degeneracy parameters in redundant calibration for MWA Phase II data.

The degeneracy projection (DP) technique introduced in our work is a process where we look for the best fit 4 degeneracy parameters for input solutions (e.g., \texttt{OMNICAL} solutions) that makes them comparable to the target solutions (e.g., FHD solutions). 

We perform DP in two cases. First, redundant calibration cannot provide a correct answer for these degeneracy parameters, necessitating an absolute calibration after \texttt{OMNICAL}. We do this by projecting \texttt{OMNICAL} solutions in the degenerate space to FHD solutions because FHD is our best guess about the sky information. The other case is when we are averaging \texttt{OMNICAL} solutions from a set of adjacent observations (here an observation refers to a single 112 second file), the degeneracy parameters may vary slightly from observation to observation. Simply averaging solutions with inconsistent degeneracy parameters can bias the average in an unknown direction. We pick one observation's solutions to be the target solutions, and project degeneracy of other observation solutions to this target before averaging them. 

\subsection{Degeneracy description}
DP for gain amplitudes is straightforward. The $\eta$'s of the input solutions are chosen to have the same average over all tiles as that of target solutions per polarization per frequency. 

As we mentioned in section 5.1, there are 3 degeneracy parameters in phase. There is actually one extra phase offset degeneracy parameter for MWA Phase II array. Since there is no inter-hexagon tile sharing the same baseline type as any intra-hexagon baseline, adding a uniform phase offset to gains of all tiles in one of the hexagons does not break any visibility redundancy. Thus the offset terms $\psi$ are treated separately for north hexagon ($\psi_{N}$) and south hexagon ($\psi_{S}$), while the phase slope $\vec{\Phi}$ is the same for both hexagons. 

To solve for these phase parameters, as mentioned in section 5.1, we can fit a plane in $(x,y,\Delta\Psi)$ space, where $\Delta\Psi$ is the phase difference between the input gain solutions and target gain solutions. 
\begin{equation}
\Delta{\Psi_i}=\Phi_xx_i+\Phi_yy_i+\psi
\label{EQdplane2}
\end{equation}
However, when we difference the phase between two complex numbers, the outcome can have a $2\pi$ ambiguity, i.e., a phase wrap. If the phase wrapping happens frequently, we are not able to directly fit a plane as Equation \ref{EQdplane2}.

\subsection{DP without phase wrapping}
\label{DPfitting}
We first consider the case where there is no phase wrapping. Practically, this case occurs in two places in our analysis. First, when we do FOcal, we are running \texttt{OMNICAL} on FHD calibrated data. If FHD were to calibrate it again, it should return ones as the solutions (i.e. no calibration needed).  Thus DP in FOcal is equivalent to projecting \texttt{OMNICAL} solutions to ones, or in other words, we do not want \texttt{OMNICAL} to add extra non-zero values to degeneracy parameters which have already been calibrated by FHD. Since in this case \texttt{OMNICAL} is looking for solutions around 1.0, the phase differences between \texttt{OMNICAL} solutions and 0.0 are small, so there is no phase wrapping. 

The second place where phase wrapping is not an issue is when comparing \texttt{OMNICAL} solutions from adjacent observations. These solutions are very similar to each other, and when we do DP between close solutions, we do not have to worry about phase wrapping. It is safe to directly apply plane fitting.

\begin{figure}
\centering
\includegraphics[width=0.8\linewidth]{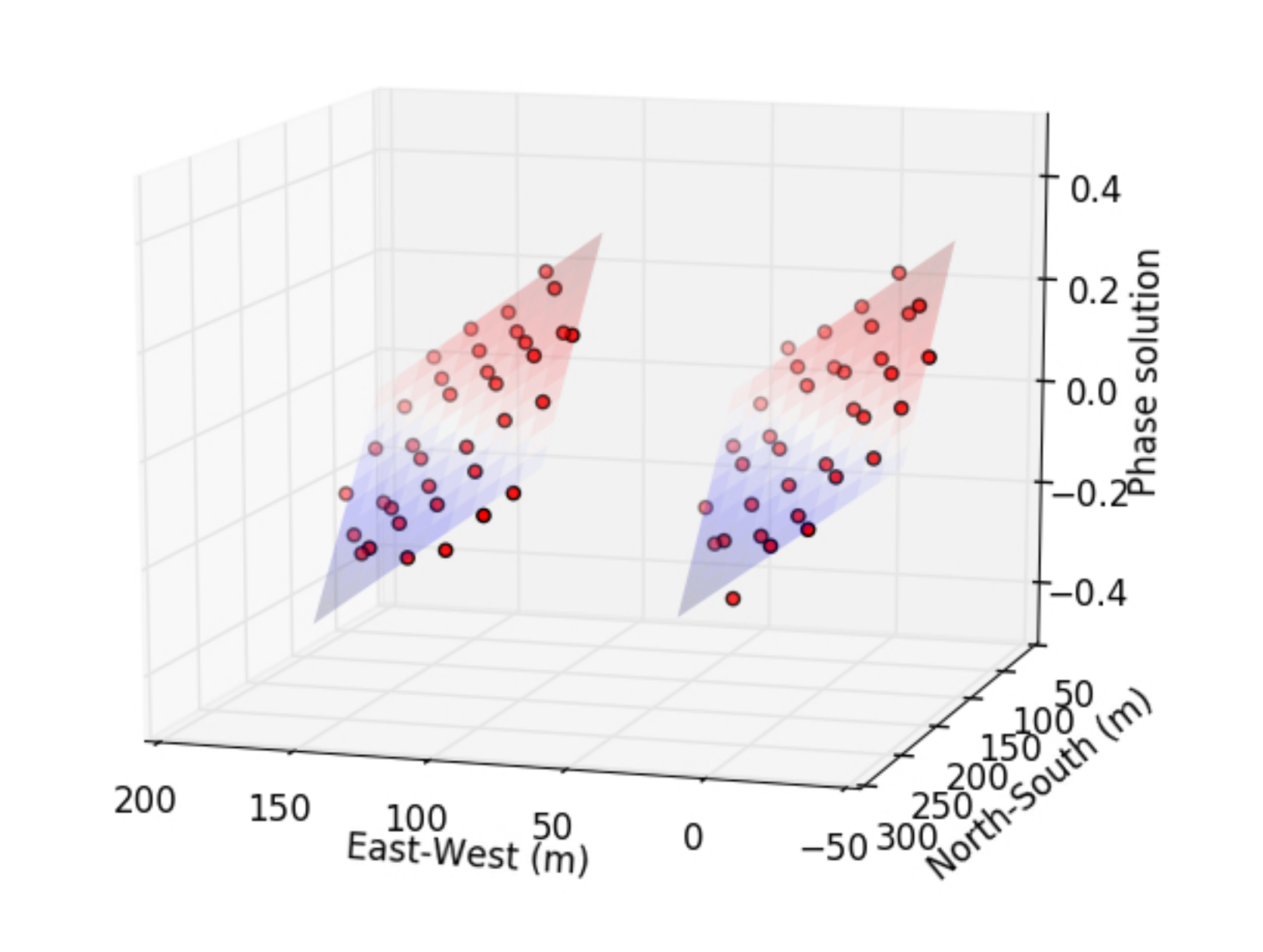}
\caption{An example of plane fitting in FOcal. Red dots represents the \texttt{OMNICAL} phase solutions (or equivalently, its difference from 0) at a single frequency single polarization versus ideal tile positions. x and y axes represents East-West positions and North-South positions, respectively, and z axis represents the phase (in radians). The two planes are fitted results. These two planes have the same $\vec{\Phi}$ but different phase offset.}
\label{FIGdegen}
\end{figure}

To calculate the best fit for Equation \ref{EQdplane2}, we minimize the quantity in Equation \ref{EQminplane}:
\begin{equation}
\begin{split}
\chi^2=&\sum_{i_N}(\Delta\Psi_{i_N}-\Phi_xx_{i_N}-\Phi_yy_{i_N}-\psi_N)^2\\
&+\sum_{i_S}(\Delta\Psi_{i_S}-\Phi_xx_{i_S}-\Phi_yy_{i_S}-\psi_S)^2,
\end{split}
\label{EQminplane}
\end{equation}
where $i_N$ is the tile index in north hexagon and $i_S$ is the tile index in south hexagon. An example of fitting result in FOcal at a single frequency single polarization is shown in Figure \ref{FIGdegen}.

\subsection{DP with phase wrapping}
Now we discuss the case where phase wrapping shows up frequently. This happens when we do OFcal. We project \texttt{OMNICAL} solutions on raw data to FHD solutions. The phase difference between \texttt{OMNICAL} and FHD is normally large. We have to unwrap the phase in 2 dimensions before plane fitting, which is challenging. Instead of directly fitting a plane, we choose to calculate a rough value of these phase parameters and remove them so that we have a close answer to our target, then apply our method in section \ref{DPfitting}.

The first step is to remove the $\psi$ terms for both hexagons by setting tile 1001 as the reference tile for north hexagon and setting tile 1072 as the reference tile for south hexagon. For each observation, the input phase solution of the reference tile is shifted to the target phase solution of that tile in the reference observation, and simultaneously the phases of all other tiles in the corresponding hexagon sub-array are shifted by the same amount. In addition, the reference tile functions as the origin for each hexagon sub-array, i.e., any tile position vector in that sub-array $\vec{r}_i$ originates from the reference tile, i.e., $\vec{r}_i=\vec{r}^{\prime}_i-\vec{r}^{\prime}_{reference}$, thus $\psi_N$ and $\psi_S$ vanish at this point. The phase degeneracy reduces to $\vec{\Phi}\cdot\vec{r}_i$.

By removing the phase offset terms for both hexagons, we only left with two degeneracy parameters in $\vec{\Phi}$. To illustrate the fitting for $\vec{\Phi}$, we select two basis vectors to describe the tile positions:
\begin{equation}
\begin{split}
\vec{a}_1&=14\hat{x}\\
\vec{a}_2&=-7\hat{x}-7\sqrt{3}\hat{y}
\end{split}
\label{EQposbasis}
\end{equation}
where $\hat{x}$ represents a vector pointing East direction with a length of 1 meter, and $\hat{y}$ represents a vector pointing to North direction with a length of 1 meter. Any tile location can be represented as:

\begin{equation}
\vec{r}_i = n_{1i}\vec{a}_1+n_{2i}\vec{a}_2
\end{equation}
Where $n_{1i}$ and $n_{2i}$ are integers (see Figure \ref{FIGpos} for tile positions). Note here $\vec{r}_i$ originates from tile 1001 for North hexagon, and tile 1072 for South hexagon. To illustrate the phase slope $\vec{\Phi}$, we use the basis shown in Equation \ref{EQphibasis}:
\begin{equation}
 \begin{split}
 \vec{b}_1=&\frac{\hat{z}\times\vec{a}_2}{\hat{z}\cdot(\vec{a}_2\times\vec{a}_1)}\\
 \vec{b}_2=&\frac{\vec{a}_1\times\hat{z}}{\hat{z}\cdot(\vec{a}_2\times\vec{a}_1)}
 \end{split}
 \label{EQphibasis}
\end{equation}
where $\hat{z}$ is a unit vector perpendicular to the plane of the array. Thus we have
\begin{equation}
\vec{a}_i\cdot\vec{b}_j=\delta_{ij}
\end{equation}
$\vec{\Phi}$ is represented as:
\begin{equation}
\vec{\Phi}=\alpha_1\vec{b}_1+\alpha_2\vec{b}_2
\end{equation}
The two components of $\vec{\Phi}$ are parameterized as $\alpha_1$ and $\alpha_2$. There is no cross term in dot product of $\vec{\Phi}$ and $\vec{r}_i$, as Equation \ref{EQdotproduct} shows.
\begin{equation}
\vec{\Phi}\cdot\vec{r}_i=\alpha_1n_{1i}+\alpha_2n_{2i}
\label{EQdotproduct}
\end{equation}
As we know unwrapping phase in two dimensions is difficult, but unwrapping phase in one dimension is easier to do. The reason we choose \ref{EQposbasis} as position basis and \ref{EQphibasis} as phase slope basis is we want to pick two non-parallel directions, which are $\vec{a}_1$ and $\vec{a}_2$ in tile position space, and do one dimensional phase unwrapping and fitting in these two directions separately. The corresponding phase slope in $\vec{a}_1$ and $\vec{a}_2$ are $\alpha_1$ and $\alpha_2$, respectively. Although there is also degeneracy for $\alpha_i$ ($\alpha_i+2\pi{N}$, where N is an integer, is also a solution), by evaluating Equation \ref{EQdotproduct}, any $2\pi$ wrap should vanish because the wrapping term gets multiplied by an integer; thus the final result is unique.

After a rough estimation of degeneracy parameters is solved in this fashion, we apply them to the input solution. At this point the input solutions and target solutions are close. To get a finer solution, we further do a plane fitting as in Section \ref{DPfitting}.

\section{\texttt{OMNICAL} convergence}
\label{convSec}
The \texttt{OMNICAL} package has shown good computational efficiency in redundant calibration. However, we have discovered a convergence issue of \texttt{lincal} in \texttt{OMNICAL}. In our work, we have done tests on convergence by using different starting points for calibration. Ideally, the solution to the least square problem in \texttt{lincal} should converge to the same answer regardless of what starting points we give it. However, \texttt{OMNICAL} only converges to a level of 0.1\% in our data set. This level of uncertainty is above the EoR signal. 

In this work, we have solved this issue and have solutions converged to machine precision. All our results presented in this paper do not have this problem.

As an example of different starting points for \texttt{OMNICAL}, we can use different baseline groups in \texttt{firstcal}, which yields the same phase slopes but different phase offset results.  This in turn leads to different results from \texttt{logcal}, thus we have different starting points for \texttt{lincal}. Not only we can use \texttt{firstcal} to get an initial guess for the phase solutions, but also we can implement a rough calibration method introduced by \citep{Zheng2016}, which we call \texttt{roughcal}. The relation between the phase of true visibilities, data, and gains is given by Equation \ref{EQphsd}. If we know the phases of true visibilities for baseline type (1001,1005) ($\theta_{1001,1005}$) and type (1001,1006) ($\theta_{1001,1006}$) (see Figure \ref{FIGpos} for information of baseline types), and $\phi_{1001}$, we are able to solve for $\phi_{1005}$ and $\phi_{1006}$. With $\phi_{1005}$ and $\phi_{1006}$ solved, we can move forward to solve for $\phi_{1002}$, $\phi_{1010}$, $\phi_{1011}$ and $\phi_{1012}$, and so on and so forth. This guarantees us to cover all tiles across this sub-array. With this incomplete information, i.e., only the data from two types of baselines, we are able to get a rough guess for phase solutions per tile. The starting point for this process is the knowledge of $\theta_{1001,1005}$, $\theta_{1001,1006}$ and $\phi_{1001}$. We are actually free to choose these three parameters because of the phase degeneracy. The most straightforward choice is $\theta_{1001,1005} = \gamma_{1001,1005}$, $\theta_{1001,1006} = \gamma_{1001,1006}$, and $\phi_{1001} = 0$, where $\gamma_{ij}$ is as defined in Equation \ref{EQphsd}.

Using either different baseline subsets for \texttt{firstcal} or using \texttt{roughcal} instead of \texttt{firstcal} has the same effect: different starting points for \texttt{OMNICAL}.
After calibration using any two sets of different starting points, we use the degeneracy projection approach to force these two sets of solutions to have the same degeneracy parameters. By comparing these two sets of results, we see a level of 0.1\% of difference in real data calibration. This level of calibration uncertainty is significant because a fraction $10^{-3}$ of foregrounds is still brighter than the EoR signal.

In the algorithmic implementation of \texttt{OMNICAL} available at \url{https://github.com/jeffzhen/omnical}, the \texttt{lincal} iteration solves the least squares problem by taking partial derivatives of $\chi^2$ given by Equation \ref{EQlinchi1} with respect to each individual parameter, forcing each partial derivative to be 0 to solve for the corresponding parameter, updating the solutions by a weighted average between the solutions from previous iteration and the new solutions. 
\begin{equation}
\label{EQlinchi1}
\chi^2 = \sum_{ij}|v_{ij} - g_i{g_j}^*y_{ij}|^2
\end{equation}
This algorithm is equivalent to iterating along the parameters axes, which is not as robust as approaching the local minimum along the real gradient in the parameter space. 

To solve for the local minimum of $\chi^2$, we add an extra step to obtain a finer convergence. We write Equation \ref{EQlincal} into a matrix form as Equation \ref{EQmatrix} \citep{Liu2010}, where \texttt{d} is a $2N_{redundant\_baselines}$ dimensional vector, \texttt{x} is a $2N_{unique\_baselines} + 2N_{tiles}$ dimensional vector, and \texttt{A} is a $2N_{redundant\_baselines} \times (2N_{unique\_baselines} + 2N_{tiles})$ dimensional matrix. 

\begin{equation}
\label{EQmatrix}
\resizebox{1\hsize}{!}{$\underbrace{\left(\begin{array}{c}
\Re(v_{ij}-g_{i}^0{g_{j}^0}^*y_{ij}^0) \\
\Im(v_{ij}-g_{i}^0{g_{j}^0}^*y_{ij}^0) \\
\vdots 
\end{array} \right)}_{\equiv \mathbf{d}}=
\underbrace{\left( \begin{array}{ccccccccc}
\Re({g_{j}^0}^*y_{ij}^0) & -\Im({g_{j}^0}^*y_{ij}^0) & \Re(g_{i}^0y_{ij}^0) & \Im(g_{i}^0y_{ij}^0) & \cdots & \Re(g_{i}^0{g_{j}^0}^*) & -\Im(g_{i}^0{g_{j}^0}^*) & \cdots\\
\Im({g_{j}^0}^*y_{ij}^0) & \Re({g_{j}^0}^*y_{ij}^0) & \Im(g_{i}^0y_{ij}^0) & -\Re(g_{i}^0y_{ij}^0) & \cdots & \Im(g_{i}^0{g_j^0}^*) & \Re(g_{i}^0{g_j^0}^*) & \cdots\\
\vdots & \vdots & \vdots & \vdots & \ddots & \vdots & \vdots & \cdots 
\end{array} \right)}_{\equiv \mathbf{A}}
\underbrace{\left(\begin{array}{c}
\Re(\Delta g_{i}) \\
\Im(\Delta g_{i}) \\
\Re(\Delta g_{j}) \\
\Im(\Delta g_{j}) \\
\vdots \\
\Re(\Delta y_{ij}) \\
\Im(\Delta y_{ij}) \\
\vdots 
\end{array} \right)}_{\equiv \mathbf{x}}$}
\end{equation}

The solution to \texttt{x} is given by Equation \ref{EQlstsol} \citep{Liu2010}:
\begin{equation}
\label{EQlstsol}
x = (A^TA)^+A^Td,
\end{equation}
where the `+' sign denotes Moore-Penrose pseudo-inverse. This step is more computationally expensive not only because the matrix \texttt{A} is large, but also it needs to be updated and find the pseudo-inverse per iteration for each time and frequency sample. In our calibration, we run \texttt{OMNICAL} first, followed by this new step to get solutions to converge. The difference in solutions from different starting points is reduced to machine precision. Although we have made this approach work within a reasonable amount of time for MWA Phase II data, this method is still not time efficient enough. For future experiments with redundant array such as Hydrogen Epoch of Reionization Array (HERA; \citep{DeBoer2017}), this matrix is much larger than that of MWA Phase II, thus higher efficiency in redundant calibration is desired. We will further explore a better approach in future work.

\bibliographystyle{aasjournal}
\bibliography{library}

\begin{thebibliography}{}
\expandafter\ifx\csname natexlab\endcsname\relax\def\natexlab#1{#1}\fi
\providecommand{\url}[1]{\href{#1}{#1}}

\bibitem[{Barry {et~al.}(2016)Barry, Hazelton, Sullivan, Morales, \&
  Pober}]{Barry2016}
Barry, N., Hazelton, B., Sullivan, I., Morales, M., \& Pober, J. 2016, Monthly
  Notices of the Royal Astronomical Society, 461, 3135

\bibitem[{Barry(2018)}]{barrythesis}
Barry, N.~A. 2018, PhD thesis, University of Washington

\bibitem[{Beardsley {et~al.}(2012)Beardsley, Hazelton, Morales, Capallo, Goeke,
  Emrich, Lonsdale, Arcus, Barnes, Bernardi, {et~al.}}]{beardsley2012new}
Beardsley, A., Hazelton, B., Morales, M.~F., {et~al.} 2012, Monthly Notices of
  the Royal Astronomical Society, 425, 1781

\bibitem[{Beardsley {et~al.}(2016)Beardsley, Hazelton, Sullivan, Carroll,
  Barry, Rahimi, Pindor, Trott, Line, Jacobs, {et~al.}}]{Beardsley2016}
Beardsley, A., Hazelton, B., Sullivan, I., {et~al.} 2016, The Astrophysical
  Journal, 833, 102

\bibitem[{Bowman {et~al.}(2009)Bowman, Morales, \&
  Hewitt}]{bowman2009foreground}
Bowman, J.~D., Morales, M.~F., \& Hewitt, J.~N. 2009, The Astrophysical
  Journal, 695, 183

\bibitem[{{Bowman} {et~al.}(2013){Bowman}, {Cairns}, {Kaplan}, {Murphy},
  {Oberoi}, {Staveley-Smith}, {Arcus}, {Barnes}, {Bernardi}, {Briggs}, {Brown},
  {Bunton}, {Burgasser}, {Cappallo}, {Chatterjee}, {Corey}, {Coster},
  {Deshpande}, {deSouza}, {Emrich}, {Erickson}, {Goeke}, {Gaensler},
  {Greenhill}, {Harvey-Smith}, {Hazelton}, {Herne}, {Hewitt},
  {Johnston-Hollitt}, {Kasper}, {Kincaid}, {Koenig}, {Kratzenberg}, {Lonsdale},
  {Lynch}, {Matthews}, {McWhirter}, {Mitchell}, {Morales}, {Morgan}, {Ord},
  {Pathikulangara}, {Prabu}, {Remillard}, {Robishaw}, {Rogers}, {Roshi},
  {Salah}, {Sault}, {Shankar}, {Srivani}, {Stevens}, {Subrahmanyan}, {Tingay},
  {Wayth}, {Waterson}, {Webster}, {Whitney}, {Williams}, {Williams}, \&
  {Wyithe}}]{bowman2013mwa}
{Bowman}, J.~D., {Cairns}, I., {Kaplan}, D.~L., {et~al.} 2013, \pasa, 30, e031

\bibitem[{Carroll {et~al.}(2016)Carroll, Line, Morales, Barry, Beardsley,
  Hazelton, Jacobs, Pober, Sullivan, Webster, {et~al.}}]{Carroll2016}
Carroll, P., Line, J., Morales, M., {et~al.} 2016, Monthly Notices of the Royal
  Astronomical Society, 461, 4151

\bibitem[{Datta {et~al.}(2010)Datta, Bowman, \& Carilli}]{Datta2010}
Datta, A., Bowman, J., \& Carilli, C. 2010, The Astrophysical Journal, 724, 526

\bibitem[{DeBoer {et~al.}(2017)DeBoer, Parsons, Aguirre, Alexander, Ali,
  Beardsley, Bernardi, Bowman, Bradley, Carilli, {et~al.}}]{DeBoer2017}
DeBoer, D.~R., Parsons, A.~R., Aguirre, J.~E., {et~al.} 2017, Publications of
  the Astronomical Society of the Pacific, 129, 045001

\bibitem[{{Dillon} \& {Parsons}(2016)}]{dillon2016hera}
{Dillon}, J.~S., \& {Parsons}, A.~R. 2016, \apj, 826, 181

\bibitem[{Dillon {et~al.}(2015)Dillon, Neben, Hewitt, Tegmark, Barry,
  Beardsley, Bowman, Briggs, Carroll, de~Oliveira-Costa,
  {et~al.}}]{dillon2015empirical}
Dillon, J.~S., Neben, A.~R., Hewitt, J.~N., {et~al.} 2015, Physical Review D,
  91, 123011

\bibitem[{{Dillon} {et~al.}(2017){Dillon}, {Kohn}, {Parsons}, {Aguirre}, {Ali},
  {Bernardi}, {Kern}, {Li}, {Liu}, {Nunhokee}, \& {Pober}}]{dillon2017redcal}
{Dillon}, J.~S., {Kohn}, S.~A., {Parsons}, A.~R., {et~al.} 2017, ArXiv
  e-prints, arXiv:1712.07212

\bibitem[{Ewall-Wice {et~al.}(2017)Ewall-Wice, Dillon, Liu, \&
  Hewitt}]{ewall2017impact}
Ewall-Wice, A., Dillon, J.~S., Liu, A., \& Hewitt, J. 2017, Monthly Notices of
  the Royal Astronomical Society, stx1221

\bibitem[{Ewall-Wice {et~al.}(2016)Ewall-Wice, Dillon, Hewitt, Loeb, Mesinger,
  Neben, Offringa, Tegmark, Barry, Beardsley, {et~al.}}]{ewall2016first}
Ewall-Wice, A., Dillon, J.~S., Hewitt, J., {et~al.} 2016, Monthly Notices of
  the Royal Astronomical Society, 460, 4320

\bibitem[{Furlanetto(2016)}]{furlanetto201621}
Furlanetto, S.~R. 2016, in Understanding the Epoch of Cosmic Reionization
  (Springer), 247--280

\bibitem[{{Furlanetto} {et~al.}(2006){Furlanetto}, {Oh}, \&
  {Briggs}}]{furlanetto2006}
{Furlanetto}, S.~R., {Oh}, S.~P., \& {Briggs}, F.~H. 2006, \physrep, 433, 181

\bibitem[{Hazelton {et~al.}(2017)Hazelton, Jacobs, Pober, \&
  Beardsley}]{JHazelton2017}
Hazelton, B.~J., Jacobs, D.~C., Pober, J.~C., \& Beardsley, A.~P. 2017, The
  Journal of Open Source Software, 2, doi:10.21105/joss.00140

\bibitem[{Hazelton {et~al.}(2013)Hazelton, Morales, \& Sullivan}]{Hazelton2013}
Hazelton, B.~J., Morales, M.~F., \& Sullivan, I.~S. 2013, The Astrophysical
  Journal, 770, 156

\bibitem[{Hurley-Walker {et~al.}(2016)Hurley-Walker, Callingham, Hancock,
  Franzen, Hindson, Kapi{\'n}ska, Morgan, Offringa, Wayth, Wu,
  {et~al.}}]{Hurley-Walker2016}
Hurley-Walker, N., Callingham, J.~R., Hancock, P.~J., {et~al.} 2016, Monthly
  Notices of the Royal Astronomical Society, 464, 1146

\bibitem[{Intema {et~al.}(2017)Intema, Jagannathan, Mooley, \&
  Frail}]{intema2017gmrt}
Intema, H., Jagannathan, P., Mooley, K., \& Frail, D. 2017, Astronomy \&
  Astrophysics, 598, A78

\bibitem[{Jacobs {et~al.}(2016)Jacobs, Hazelton, Trott, Dillon, Pindor,
  Sullivan, Pober, Barry, Beardsley, Bernardi, {et~al.}}]{Jacobs2016}
Jacobs, D.~C., Hazelton, B., Trott, C., {et~al.} 2016, The Astrophysical
  Journal, 825, 114

\bibitem[{{Line} {et~al.}(in~prep.)}]{ORBCommLine}
{Line}, {et~al.} in~prep.

\bibitem[{Liu {et~al.}(2014)Liu, Parsons, \& Trott}]{Liu2014}
Liu, A., Parsons, A.~R., \& Trott, C.~M. 2014, Physical Review D, 90, 023018

\bibitem[{Liu {et~al.}(2010)Liu, Tegmark, Morrison, Lutomirski, \&
  Zaldarriaga}]{Liu2010}
Liu, A., Tegmark, M., Morrison, S., Lutomirski, A., \& Zaldarriaga, M. 2010,
  Monthly Notices of the Royal Astronomical Society, 408, 1029

\bibitem[{{Mesinger} {et~al.}(2011){Mesinger}, {Furlanetto}, \&
  {Cen}}]{mesinger2011}
{Mesinger}, A., {Furlanetto}, S., \& {Cen}, R. 2011, \mnras, 411, 955

\bibitem[{Morales {et~al.}(2012)Morales, Hazelton, Sullivan, \&
  Beardsley}]{Morales2012}
Morales, M.~F., Hazelton, B., Sullivan, I., \& Beardsley, A. 2012, The
  Astrophysical Journal, 752, 137

\bibitem[{Morales \& Wyithe(2010)}]{morales2010reionization}
Morales, M.~F., \& Wyithe, J. S.~B. 2010, Annual review of astronomy and
  astrophysics, 48, 127

\bibitem[{Neben {et~al.}(2015)Neben, Bradley, Hewitt, Bernardi, Bowman, Briggs,
  Cappallo, Deshpande, Goeke, Greenhill, {et~al.}}]{neben2015measuring}
Neben, A., Bradley, R., Hewitt, J.~N., {et~al.} 2015, Radio Science, 50, 614

\bibitem[{Neben {et~al.}(2016)Neben, Bradley, Hewitt, DeBoer, Parsons, Aguirre,
  Ali, Cheng, Ewall-Wice, Patra, {et~al.}}]{neben2016hydrogen}
Neben, A.~R., Bradley, R.~F., Hewitt, J.~N., {et~al.} 2016, The Astrophysical
  Journal, 826, 199

\bibitem[{{Nikolic} {et~al.}(2017){Nikolic}, {Carilli}, \& {HERA
  Collaboration}}]{nikolic2017hera}
{Nikolic}, B., {Carilli}, C., \& {HERA Collaboration}. 2017, ArXiv e-prints,
  arXiv:1709.05245

\bibitem[{Noorishad {et~al.}(2012)Noorishad, Wijnholds, van Ardenne, \& Van
  Der~Hulst}]{noorishad2012redundancy}
Noorishad, P., Wijnholds, S.~J., van Ardenne, A., \& Van Der~Hulst, J. 2012,
  Astronomy \& Astrophysics, 545, A108

\bibitem[{Offringa {et~al.}(2015)Offringa, Wayth, Hurley-Walker, Kaplan, Barry,
  Beardsley, Bell, Bernardi, Bowman, Briggs, {et~al.}}]{Offringa2015}
Offringa, A., Wayth, R., Hurley-Walker, N., {et~al.} 2015, Publications of the
  Astronomical Society of Australia, 32

\bibitem[{Parsons {et~al.}(2012{\natexlab{a}})Parsons, Pober, McQuinn, Jacobs,
  \& Aguirre}]{parsons2012sensitivity}
Parsons, A., Pober, J., McQuinn, M., Jacobs, D., \& Aguirre, J.
  2012{\natexlab{a}}, The Astrophysical Journal, 753, 81

\bibitem[{Parsons \& Backer(2009)}]{Parsons2009}
Parsons, A.~R., \& Backer, D.~C. 2009, The Astronomical Journal, 138, 219

\bibitem[{Parsons {et~al.}(2012{\natexlab{b}})Parsons, Pober, Aguirre, Carilli,
  Jacobs, \& Moore}]{Parsons2012}
Parsons, A.~R., Pober, J.~C., Aguirre, J.~E., {et~al.} 2012{\natexlab{b}}, The
  Astrophysical Journal, 756, 165

\bibitem[{Patil {et~al.}(2016)Patil, Yatawatta, Zaroubi, Koopmans, de~Bruyn,
  Jeli{\'c}, Ciardi, Iliev, Mevius, Pandey, {et~al.}}]{patil2016systematic}
Patil, A.~H., Yatawatta, S., Zaroubi, S., {et~al.} 2016, Monthly Notices of the
  Royal Astronomical Society, 463, 4317

\bibitem[{Pober {et~al.}(2013)Pober, Parsons, Aguirre, Ali, Bradley, Carilli,
  DeBoer, Dexter, Gugliucci, Jacobs, {et~al.}}]{Pober2013}
Pober, J.~C., Parsons, A.~R., Aguirre, J.~E., {et~al.} 2013, The Astrophysical
  Journal Letters, 768, L36

\bibitem[{Procopio {et~al.}(2017)Procopio, Wayth, Line, Trott, Intema,
  Mitchell, Pindor, Riding, Tingay, Bell, {et~al.}}]{procopio2017high}
Procopio, P., Wayth, R., Line, J., {et~al.} 2017, Publications of the
  Astronomical Society of Australia, 34

\bibitem[{Salvini \& Wijnholds(2014)}]{Salvini2014}
Salvini, S., \& Wijnholds, S.~J. 2014, Astronomy \& Astrophysics, 571, A97

\bibitem[{Sullivan {et~al.}(2012)Sullivan, Morales, Hazelton, Arcus, Barnes,
  Bernardi, Briggs, Bowman, Bunton, Cappallo, {et~al.}}]{Sullivan2012}
Sullivan, I., Morales, M.~F., Hazelton, B., {et~al.} 2012, The Astrophysical
  Journal, 759, 17

\bibitem[{Thyagarajan {et~al.}(2013)Thyagarajan, Shankar, Subrahmanyan, Arcus,
  Bernardi, Bowman, Briggs, Bunton, Cappallo, Corey,
  {et~al.}}]{Thyagarajan2013}
Thyagarajan, N., Shankar, N.~U., Subrahmanyan, R., {et~al.} 2013, The
  Astrophysical Journal, 776, 6

\bibitem[{Thyagarajan {et~al.}(2015)Thyagarajan, Jacobs, Bowman, Barry,
  Beardsley, Bernardi, Briggs, Cappallo, Carroll, Corey,
  {et~al.}}]{Thyagarajan2015}
Thyagarajan, N., Jacobs, D.~C., Bowman, J.~D., {et~al.} 2015, The Astrophysical
  Journal, 804, 14

\bibitem[{Tingay {et~al.}(2013)Tingay, Goeke, Bowman, Emrich, Ord, Mitchell,
  Morales, Booler, Crosse, Wayth, {et~al.}}]{Tingay2013}
Tingay, S., Goeke, R., Bowman, J.~D., {et~al.} 2013, Publications of the
  Astronomical Society of Australia, 30

\bibitem[{Trott \& Wayth(2016)}]{trott2016spectral}
Trott, C.~M., \& Wayth, R.~B. 2016, Publications of the Astronomical Society of
  Australia, 33

\bibitem[{Trott {et~al.}(2012)Trott, Wayth, \& Tingay}]{Trott2012}
Trott, C.~M., Wayth, R.~B., \& Tingay, S.~J. 2012, The Astrophysical Journal,
  757, 101

\bibitem[{Trott {et~al.}(2016)Trott, Pindor, Procopio, Wayth, Mitchell,
  McKinley, Tingay, Barry, Beardsley, Bernardi, {et~al.}}]{trott2016chips}
Trott, C.~M., Pindor, B., Procopio, P., {et~al.} 2016, The Astrophysical
  Journal, 818, 139

\bibitem[{Vedantham {et~al.}(2012)Vedantham, Shankar, \&
  Subrahmanyan}]{Vedantham2012}
Vedantham, H., Shankar, N.~U., \& Subrahmanyan, R. 2012, The Astrophysical
  Journal, 745, 176

\bibitem[{{Wayth} {et~al.}(in~prep.)}]{MWAPhaseII}
{Wayth}, R.~B., {et~al.} in~prep.

\bibitem[{Wieringa(1992)}]{Wieringa1992}
Wieringa, M.~H. 1992, Exp. Astron., 2, 203

\bibitem[{Zheng {et~al.}(2014)Zheng, Tegmark, Buza, Dillon, Gharibyan, Hickish,
  Kunz, Liu, Losh, Lutomirski, {et~al.}}]{Zheng2014}
Zheng, H., Tegmark, M., Buza, V., {et~al.} 2014, Monthly Notices of the Royal
  Astronomical Society, 445, 1084

\bibitem[{Zheng {et~al.}(2016)Zheng, Tegmark, Dillon, Liu, Neben, Tribiano,
  Bradley, Buza, Ewall-Wice, Gharibyan, {et~al.}}]{Zheng2016}
Zheng, H., Tegmark, M., Dillon, J., {et~al.} 2016, Monthly Notices of the Royal
  Astronomical Society, stw2910

\end{thebibliography}

\end{document}